\documentclass[12pt,journal,a4paper,onecolumn,draftcls]{IEEEtran}

\usepackage{cite}
\usepackage{pgfplots}
\usetikzlibrary{spy}
\usepackage[caption=false,font=footnotesize]{subfig}
\usepackage{mymacros}
\usepgfplotslibrary{external}
{cell-free-system-info} 
\tikzsetexternalprefix{figures/}

\newcommand{\changed}[1]{{\color{black}#1}}

\tikzset{every picture/.style=thick}

\newtheorem{theorem}{Theorem}
\newtheorem{corollary}[theorem]{Corollary}

\title{Techniques for System Information Broadcast in Cell-Free Massive MIMO}

\author{Marcus Karlsson, Emil Bj{\"o}rnson and Erik G. Larsson\thanks{\copyright 2018 IEEE. Personal use of this material is permitted. Permission from IEEE must be obtained for all other uses, in any current or future media, including reprinting/republishing this material for advertising or promotional purposes, creating new collective works, for resale or redistribution to servers or lists, or reuse of any copyrighted component of this work in other works.}
\thanks{The authors are with the Department of Electrical Engineering (ISY), Link{\"o}ping University, 581 83 Link{\"o}ping, Sweden (email: \{marcus.karlsson, emil.bjornson,  erik.g.larsson\}@liu.se).} \thanks{This work was supported in part by the Swedish Research Council (VR), and ELLIIT.}}

\IEEEoverridecommandlockouts
\begin{document}
\maketitle
\begin{abstract}
	\changed{We consider transmission of system information in a cell-free massive \MIMO system, when the transmitting access points do not have any channel state information and the receiving terminal has to estimate the channel based on downlink pilots. We analyze the system performance in terms of outage rate and coverage probability, and use space-time block codes to increase performance. We propose a heuristic method for pilot/data power optimization that can be applied without any channel state information at the access points. We also analyze the problem of grouping the access points, which is needed when the single-antenna access points jointly transmit a space-time block code.}
\end{abstract}

\section{Introduction}

\begin{figure}
	\centering
	\usetikzlibrary{calc}
\begin{tikzpicture} [connection/.style={dashed,thick},
mobile/.pic = {%
		\draw[thick] (0,0) rectangle (1,1.61);
		\draw[thick] (1.0,1.61) -- (1.0,2.3);},
AP/.pic = {%
		\coordinate (-center) at (0,0);
		\coordinate (-top) at (0,1);
		\coordinate (-left-leg) at (-2,-5);
		\coordinate (-right-leg) at (2,-5);
		\draw[thick] (-left-leg) -- (-center) -- (-right-leg);
		\draw[thick] (-center) -- (-top);
		\coordinate (-left) at ($(-left-leg)!0.5!(-top)$);
		\coordinate (-right) at ($(-right-leg)!0.5!(-top)$);
		\coordinate (-bottom) at ($(-right-leg)!0.5!(-left-leg)$);
		\draw[thick,line join=round] (-top) -- +(-0.5,0.5) -- +(0.5,0.5) -- cycle;
		\draw[thick,line join=bevel] ($(-left-leg)!0.3!(-center)$) -- ($(-right-leg)!0.3!(-center)$) -- ($(-left-leg)!0.65!(-center)$) -- ($(-right-leg)!0.65!(-center)$) -- cycle;},
]
\node[draw,rectangle] (cpu) at (0,0) {\CPU};
\pic[scale=0.2] (one) at (0:3) {AP};
\pic[scale=0.2] (two) at (30:5) {AP};
\pic[scale=0.2] (three) at (83:4) {AP};
\pic[scale=0.2] (four) at (105:3) {AP};
\pic[scale=0.2] (five) at (130:5) {AP};
\pic[scale=0.2] (six) at (145:3) {AP};
\pic[scale=0.2] (seven) at (189:4) {AP};
\pic[scale=0.2] (eight) at (210:5) {AP};
\pic[scale=0.2] (nine) at (240:2) {AP};
\pic[scale=0.2] (ten) at (265:3) {AP};
\pic[scale=0.2] (eleven) at (298:3) {AP};
\pic[scale=0.2] (twelve) at (330:4) {AP};
\draw[connection] (cpu.east) to[out=10,in=160] (one-left);
\draw[connection] (cpu.east) to[out=30,in=160] (two-left);
\draw[connection] (cpu.north) to[out=70,in=260] (three-bottom);
\draw[connection] (cpu.north) to[out=90,in=290] (four-bottom);
\draw[connection] (cpu.north) to[out=100,in=310] (five-bottom);
\draw[connection] (cpu.west) to[out=180,in=0] (six-right);
\draw[connection] (cpu.west) to[out=180,in=0] (seven-right);
\draw[connection] (cpu.west) to[out=180,in=0] (eight-right);
\draw[connection] (cpu.south) to[out=270,in=0] (nine-right);
\draw[connection] (cpu.south) to[out=270,in=30] (ten-right);
\draw[connection] (cpu.south) to[out=270,in=110] (eleven-left);
\draw[connection] (cpu.east) to[out=0,in=180] (twelve-left);
\pic[scale=0.2] at (15:3.5) {mobile};
\pic[scale=0.2] at (45:4.5) {mobile};
\pic[scale=0.2] at (45:2.5) {mobile};
\pic[scale=0.2] at (110:4.0) {mobile};
\pic[scale=0.2] at (140:1.0) {mobile};
\pic[scale=0.2] at (160:4.0) {mobile};
\pic[scale=0.2] at (180:3.0) {mobile};
\pic[scale=0.2] at (210:3.5) {mobile};
\pic[scale=0.2] at (230:4.5) {mobile};
\pic[scale=0.2] at (230:1.5) {mobile};
\pic[scale=0.2] at (250:3.5) {mobile};
\pic[scale=0.2] at (290:1.5) {mobile};
\pic[scale=0.2] at (305:3.5) {mobile};
\pic[scale=0.2] at (335:4.5) {mobile};
\node[below] at (45:4.5) {terminal};
\node[below] at (twelve-bottom) {access point};
\end{tikzpicture}

	\caption{A cell-free massive \MIMO system where all the access points are connected to a central processing unit (\CPU). All access points serves all terminals, in the same time-frequency interval.}\label{fig:cell:free:massive:mimo}
\end{figure}
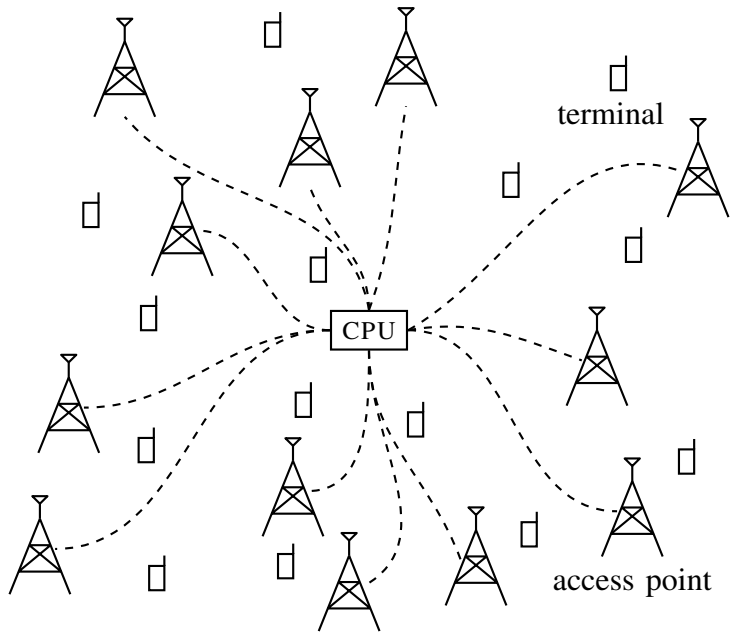

\changed{A cell-free massive multiple-input multiple-output (\MIMO) system \cite{Nayebi15,Ngo17a, Nayebi17,Nguyen17,Ngo18,Interdonato18, Bashar18}, consists of a large number of access points (\APs) distributed in an area serving all users in a coordinated fashion, using the same time-frequency resource. This is illustrated in \Figref{fig:cell:free:massive:mimo}. As in conventional (cellular) massive \MIMO systems \cite{Marzetta16,Bjornson17}, operation requires time-division duplex (\TDD) mode to be fully scalable, as channel reciprocity is used to estimate the uplink and downlink channels with uplink pilots.}

\changed{Conceptually, cell-free massive \MIMO is the same as network \MIMO (also known as coordinated multipoint with joint transmission \cite{Irmer11}), building on the principles of \cite{Shamai01}. The full-scale version of network \MIMO, when all \APs share all available information is practically infeasible \cite{Gesbert10} but efforts have been made to reap some of the benefits of network \MIMO, \cite{Zhou03}, while not sharing all available information. In \cite{Bjornson10}, for example, the \APs share the data but not the channel state information (\CSI) and in \cite{Buzzi17} only a subset of the \APs serve a particular user.}

\changed{In any practical implementation, the performance of network \MIMO is limited as the system will still suffer from interference \cite{Lozano13}, partly from the imperfect channel estimation. In much prior work, however, the impact of channel uncertainty and the cost of channel estimation have been neglected \cite{Gesbert10}. Cell-free massive \MIMO methodology, on the other hand, can quantify the cost of channel estimation and the impact of this estimation when \APs do not share \CSI. In short (paraphrasing \cite{Ngo17a}): cell-free massive \MIMO is to network \MIMO (or distributed antenna systems) what massive \MIMO is to multi-user \MIMO. While cell-free massive \MIMO can utilize the same transceiver processing as in cellular massive \MIMO for transmission and estimation, the resource allocation is fundamentally different: scheduling, power control, random access, and system information broadcast must be implemented in a distributed fashion without breaking these tasks down into separate per-cell tasks \cite{Interdonato18}. Moreover, cell-free massive \MIMO may not be able to rely on channel hardening and favorable propagation to the same extent as cellular massive \MIMO \cite{Chen18}.}

\changed{In this paper we are concerned with distributing \emph{system information} in the downlink, which is necessary for an \emph{inactive} terminal to connect to and function within the network \cite{Dahlman11,Karlsson18}. Normally when analyzing a massive \MIMO system, transmission starts with the terminal transmitting uplink pilots, in order for the \APs to estimate the (reciprocal) channel between the \APs and the terminal. However, only \emph{active} terminals, who have successfully decoded the system information, knows when and how to transmit pilots. The system information is transmitted and received without any prior \CSI. In general, this open-loop transmission---when the \APs do not have any \CSI---can be a limiting factor when it comes to network coverage, as coherent beamforming cannot be utilized. Downlink broadcasting of system information in \LTE is described in general terms in \cite{Dahlman11}; in \cite{Meng16, Karlsson18}, the downlink broadcasting of system information in a conventional, cellular massive \MIMO system was analyzed.}

Previous studies on the outage probability of wireless networks, e.g. \cite{Andrews11,Lu15,ElSawy17}, differ from the current paper in at least three aspects: First, they (sometimes implicitly) assume perfect \CSI at the receiver while we consider \CSI obtained from downlink pilots. Second, the terminals considered here are inactive; hence, no \CSI is available at the transmitter. Third, here, a terminal is served jointly by all the \APs, implying that there is no inter-cell interference. Although coverage, coverage probability, and outage probability have been mentioned in previous work on cell-free massive \MIMO \cite{Nayebi15,Ngo17a}, these papers also consider active terminals while we consider inactive ones.

This paper aims to quantify the coverage in a cell-free massive \MIMO system in terms of outage rate and coverage probability for the transmission of system information to inactive users. We analyze the coverage when the \APs as well as the terminals does not have any \CSI prior to transmission. Moreover, the effects of adding spatial diversity in terms of a space-time block code are also investigated.

\subsection{Notation}

Scalars are denoted by lower-case letters ($x$), vectors by lower-case, bold-faced letters ($\bx$), and matrices with upper-case, bold-faced letters ($\bX$). $\CN(\bzero,\bX)$ is a circularly-symmetric, complex, Gaussian random variable with zero mean and covariance matrix $\bX$. An exponentially distributed random variable with mean $\lambda^{-1}$ is denoted by $\ExpDist{\lambda}$. A chi-squared distributed random variable with $d$ degrees of freedom is denoted by $\chi^{2}_{d}$. $\real{\cdot}$ and $\imag{\cdot}$ denotes the real and imaginary part, respectively. The imaginary unit is denoted by $\iu$.

\section{System Model}\label{sec:system:model}

In many cases, it is reasonable to assume that the downlink data transmission takes place over a relatively large bandwidth (say $20$ MHz), and perhaps over a few tens of milliseconds. This allows for the data to be coded over several coherence intervals so that each codeword sees many independent channel realizations. When coding over many channel realizations, one can talk about ergodic rates; however, the downlink transmission of system information studied in this paper is considered to be on a narrow-band channel, perhaps a few $100$ kHz and with relatively low latency; thus very little frequency and time diversity is available. The transmission will take place over a single coherence interval.

The system information must always be available for a terminal wanting to connect to the network. Therefore, it must be statically allocated to the time-frequency grid. Consequently, the system information should contain as few bits as possible so as few resources as possible must be dedicated to system information. This is why coding over several coherence intervals is not considered in this paper. In essence, we want to see how much we can transmit with the minimum amount of resources.

We consider the downlink of a cell-free massive \MIMO system, aiming to convey system information to an arbitrary single-antenna terminal over a narrow-band, frequency-flat channel. The single-antenna access points (\APs) are coordinated and synchronized but lack \CSI to the terminal and resort to open-loop transmission. In order to increase reliability and coverage, the \APs may cooperate to jointly transmit an orthogonal space-time block code (\ostbc). To this end, the \APs are divided into $\Ng$ disjoint groups, each group transmitting a separate part of the \ostbc. How the groups are formed is an interesting problem in itself. For now, we consider the \APs to be grouped in an arbitrary way and leave the discussion regarding \emph{how} to group the \APs for \Secref{sec:other:colorings}.

The system consists of $\Nap$ single-antenna \APs which transmit simultaneously with the same normalized transmit power $\rho$. The aggregated signal from the $\Nap$ sources received at the terminal can be written as
\begin{equation}\label{eq:received:signal:terminal}
y = \sqrt{\rho}\sum_{m=1}^{\Nap} g_m\beta_m^{1/2}q_m + w,
\end{equation}
where $g_m$ and $\beta_m$ model the small-scale fading and the large-scale fading, respectively. $q_m$ is the symbol transmitted from \AP $m$ and $w$ is additive noise. We assume $w$ and $g_m$ are $\CN(0,1)$ and mutually independent for all $m$. We assume a quasi-static channel, where the small-scale fading is static for a coherence interval, consisting of $\tc$ symbols, and then takes a new, independent value in the next coherence interval. The large-scale fading, $\beta_m$, depends on the position of the terminal relative to the \AP and may be constant over multiple coherence intervals.

Note that, the received signal in \Eqref{eq:received:signal:terminal} only consists of useful signals and noise---no interference. This comes from the implicit assumption that the transmission of system information takes place in a dedicated time-frequency resource, as the master-information block in \LTE \cite{Dahlman11}. Because there is no interference, all \APs are assumed to transmit with full power ($\rho$). Moreover, the useful, received \emph{signal} power takes the form of a shot noise process \cite{Weber12}. This makes the setting in the current paper fundamentally different from others looking at coverage, where the \emph{interference} power is a shot noise process \cite{Weber12,Lu15}.

Now, divide the $\Nap$ \APs into $\Ng$ disjoint groups: $\mathcal{G}_1,\dots, \mathcal{G}_{\Ng}$, where all \APs in group $k$ transmit the same symbol, denoted by $x_k$. In other words, $q_m=x_k$ if $m\in\mathcal{G}_{k} $. The received signal in \Eqref{eq:received:signal:terminal} can then be written as
\begin{equation}\label{eq:received:signal:grouped}
y = \sqrt{\rho}\sum\limits_{k=1}^{\Ng}\sum_{m\in\mathcal{G}_k} g_m\beta_m^{1/2}x_k + w.
\end{equation}
By defining the \emph{effective} channel as 
\[ \bh \triangleq \left[\sum_{m\in\mathcal{G}_1} g_m\beta_m^{1/2}, \dots, \sum_{m\in\mathcal{G}_{\Ng}} g_m\beta_m^{1/2}\right]\trans\in \mC^{\Ng\times 1}, \]
and $\bx = [x_1,\dots, x_{\Ng}]\trans\in\mC^{\Ng\times 1}$, we can write \Eqref{eq:received:signal:grouped} as
\[ y = \sqrt{\rho}\bx\trans\bh + w. \]
Over $\tau$ channel uses, the terminal receives
\begin{equation}\label{eq:received:signal:vector}
\by = \sqrt{\rho}\bX\bh + \bw \in \mC^{\tau\times 1,},
\end{equation}
where 
\[ \bX \triangleq 
\left[\begin{array}{c}
\bx_1\trans\\
\bx_2\trans\\
\vdots\\
\bx_{\tau}\trans
\end{array}\right]\in\mC^{\tau\times\Ng} \]
is a matrix of collectively transmitted samples and \[ \bw \triangleq [w_1,\dots, w_\tau]\trans\sim\CN(\bzero,\bI_{\tau}) \]
is the noise vector. The effective channel, $\bh$, is an $\Ng$-dimensional zero-mean, circularly-symmetric, complex, Gaussian vector, with a diagonal covariance matrix denoted by $\bC_{\bh}$. The $k$th diagonal element of $\bC_{\bh}$ is given by 
\begin{equation}\label{eq:beta:bar}
\bar{\beta}_k \triangleq \sum\limits_{m\in\mathcal{G}_k}\beta_m.
\end{equation}

\section{Space-Time Block Codes}

A linear space-time block code (\stbc) maps a set of $\Ns$ symbols $s_1,\dots,s_{\Ns}$ onto a code matrix $\bX$, according to \cite{Larsson03}
\begin{equation}\label{eq:stbc}
\bX = \sum\limits_{n=1}^{\Ns}\An\rsn + \iu\Bn\isn,
\end{equation}
where the symbol $\sn$ is split in its real ($\rsn$) and imaginary ($\isn$) parts. The matrices $\An$ and $\Bn$ define the particular code and are complex in general. A special case of a linear \stbc is an \emph{orthogonal} \stbc (\ostbc), which has the additional property that for any $\{\sn\}$ \[ \bX\herm\bX = \bI_{\Ng}\sum\limits_{n=1}^{\Ns}|\sn|^{2}. \]
We let all symbols have the same energy $\Es\triangleq \Exp{|\sn|^{2}}$, which implies
\[ \Exp{\bX\herm\bX} = \Ns\Es\bI_{\Ng}. \]

Some of the aspects and properties of \ostbcs that are relevant for transmitting system information were covered in \cite{Karlsson18}, and will not be repeated here. One additional useful property comes from the definition of the code matrix $\bX$ in \Eqref{eq:stbc}:
\begin{equation}\label{eq:ostbcidentity:expectation:Xs}
\Exp{\bX\rsn} = \dfrac{\Es}{2}\An \text{ and } \Exp{\bX\isn} = \iu\dfrac{\Es}{2}\Bn,
\end{equation}
since the symbols are assumed to be mutually independent and the real and imaginary parts of each symbol are uncorrelated and have zero mean.

\section{Received SNR at the Terminal}\label{sec:snr}

In this section, we derive the distribution of the received \SNR at the terminal conditioned on the large-scale fading, when the \APs are jointly transmitting an \ostbc, as described in \Secref{sec:system:model}. Recall that the channel statistics depend on the large-scale fading and are therefore random when considering a randomly located terminal. The large-scale fading coefficients are, however, independent of the other randomness considered, such as the transmitted symbols, receiver noise, and small-scale fading. Thus, in this section, we consider a fixed (deterministic) set of large-scale coefficients, and derive expressions for the \SNR. 

For each realization of the effective channel $\bh$, we let $\hhat$ quantify what the receiver (terminal) knows about the effective channel. Letting $\hat{s}_n$ denote the received (possibly processed) symbol, the rate of communication over this channel, measured in bit per channel use (\bpcu), is
\[ \log(1 + \snr), \]
where the \SNR at the terminal is defined as
\begin{equation}\label{eq:LMMSE:bound}
\snr \triangleq\dfrac{\left|\Exp{\hat{s}_n\sn\conj\given \hhat}\right|^{2}}{\Es\Exp{|\hat{s}_n|^{2}\given\hhat} - \left|\Exp{\hat{s}_n\sn\conj\given\hhat}\right|^{2}}
\end{equation}
and $s_n$ is the information bearing symbol. The formula \Eqref{eq:LMMSE:bound} can be derived using results from \cite{Medard00}. Note that the \SNR in \Eqref{eq:LMMSE:bound} is random, since it is a function of the random variable $\hhat$.

\subsection{Perfect CSI}\label{sec:snr:perfect:csi}

First, we consider the case when the terminal knows the effective channel, $\bh$, perfectly. This will never happen in practice; however, if training is inexpensive we can estimate the channel with very high accuracy and essentially eliminate the estimation errors.

At the terminal, a maximum likelihood detector is used to decode the symbols. With perfect \CSI and an \ostbc, the detection of the different symbols decouples. Moreover, the real and imaginary parts of the received signal can be detected separately \cite[Section 7.4]{Larsson03}. From \cite{Karlsson18} and more generally from \cite{Larsson03}, we can write the processed real part of the symbol as (cf. \Eqref{eq:received:signal:vector})
\[ \hat{\bar{s}}_n \triangleq \real{\bh\herm\An\herm\by} = \sqrt{\rho}\norm{\bh}^{2}\rsn + \bar{z}, \]
where $\bar{z}\triangleq \real{\bh\herm\An\herm\bw}$ and
\[ \Exp{\bar{z}^{2}\given\bh} = \dfrac{1}{2} \norm{\bh}^{2}.\]
Similarly, for the imaginary part,
\[ \hat{\tilde{s}}_n \triangleq \imag{\bh\herm\Bn\herm\by} = \sqrt{\rho}\norm{\bh}^{2}\isn + \tilde{z}, \]
where $\tilde{z}\triangleq \imag{\bh\herm\Bn\herm\bw}$ and
\[ \Exp{\tilde{z}^{2}\given\bh} = \dfrac{1}{2} \norm{\bh}^{2}.\]
With $z = \bar{z} + \iu\tilde{z}$, the processed signal can now be written as
\[ \hat{s}_n \triangleq \hat{\bar{s}}_n + \iu\hat{\tilde{s}}_n =  \sqrt{\rho}\norm{\bh}^{2}\sn + z. \]
With perfect \CSI, the terminal knows $\bh$ (i.e. $\hhat= \bh$ in \Eqref{eq:LMMSE:bound}) and the noise term $z$ is uncorrelated with $\sn$. The \SNR for the received signal is given by 
\[ \snr^{\textsc{p}} \triangleq \rho\Es\norm{\bh}^{2}. \]
Each element in the effective channel is a zero-mean, circularly-symmetric, Gaussian variable; thus, the \SNR can be written as
\begin{equation}\label{eq:snr:distribution:perfect}
\snr^{\textsc{p}}\sim \dfrac{\rho\Es}{2} \sum\limits_{n=1}^{\Ng}\bar{\beta}_n\chi_{2}^{2}
\sim \sum\limits_{n=1}^{\Ng}\ExpDist{\lambda^{\textsc{p}}_n},
\end{equation}
with
\begin{equation}\label{eq:lambda:perfect}
\lambda^{\textsc{p}}_n \triangleq \dfrac{1}{\rho\Es\bar{\beta}_n}.
\end{equation}

\subsection{Imperfect CSI}\label{sec:snr:estimated:csi:general}

\begin{figure}
	\centering
\begin{tikzpicture}
\tikzstyle{frameStyle} = [draw,rectangle, text centered, minimum height = 1 cm];
\node[frameStyle,minimum width=2cm] (pilots) at (0, 0) {downlink pilots};
\node[above] at (pilots.north) {$\tp$};
\node[frameStyle, anchor=west, minimum width=5cm] (data) at (pilots.east) {system information};
\node[above] at (data.north) {$\tc - \tp$};
\end{tikzpicture}
	\caption{The downlink transmission of system information when neither the \APs nor the terminal have a priori \CSI requires downlink pilots for the terminal to estimate the channel and decode the system information. This transmission takes place over one coherence interval, consisting of $\tc$ samples, and $\tp$ samples are spent on pilots.}\label{fig:pilot:data:protocol}
\end{figure}
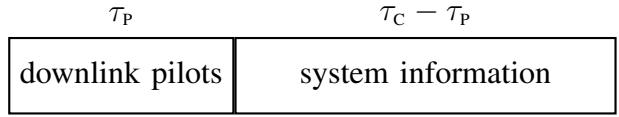

In a more practical scenario, the terminal has to estimate the channel from downlink pilots in order to detect the transmitted symbols in the \ostbc. This is illustrated in \Figref{fig:pilot:data:protocol}. The terminal uses a least-squares (\LS) estimate of the channel in favor of the commonly used minimum mean-squared error (\MMSE) estimate, since the terminal has no knowledge of the channel statistics, which is required to do \MMSE estimation. The main effect of this, at least analytically, is that the channel estimate and the estimation error are not independent, as they would be if the \MMSE estimate were used. As a result, the derivations for an achievable \SNR are a bit more tedious than in the \MMSE case, but necessary to obtain realistic results. For some special cases, it has been shown that the performance of the two estimators is identical \cite{Karlsson18}; however, for the general case, it is still unclear how the performance between \LS and \MMSE differs in this setting.

\subsubsection{Pilot Phase}

Each \AP group transmits its own pilot sequence of $\tp$ symbols. The pilot sequences are mutually orthogonal and transmitted in the pilot block $\Xp\in\mC^{\tp\times\Ng}$, satisfying
\begin{equation}\label{eq:pilot:block:property}
\Xp\herm\Xp = \tp\bI_{\Ng}.
\end{equation}
The normalization of the pilot block in \Eqref{eq:pilot:block:property} implies that the total energy spent on pilots scales with $\Nap$.

Assuming that all \APs transmit pilots with power $\pp$, the \LS channel estimate at the terminal is given by 
\[ \hhat \triangleq \left(\sqrt{\pp}\Xp\herm\Xp\right)\inv\Xp\herm\by = \bh + \be, \]
where $\by$ is given by \Eqref{eq:received:signal:vector},
\[ \be \triangleq \dfrac{1}{\sqrt{\pp}\tp}\Xp\herm\bw \Rightarrow \be\sim \CN\left(\bzero, \Ce\right),  \]
and 
\[ \Ce \triangleq \dfrac{1}{\pp\tp}\bI_{\Ng}. \]
It is important to note that, since the error $\be$ and the estimate $\hhat$ are correlated, conditioning on $\hhat$ will alter the distribution of $\be$. To be precise,
\begin{equation}\label{eq:error:conditional:distribution}
\be | \hhat \sim \CN\left(\Ucond\hhat,\Ccond\right),
\end{equation}
where
\[ \Ucond \triangleq \Ce\left(\Ce + \Ch\right)\inv \]
and 
\[ \Ccond \triangleq \left(\Ce\inv + \Ch\inv\right)\inv. \]
This is an application of \cite[Theorem 10.2]{Kay93}, shown explicitly in \cite[Lemma B.17]{Bjornson17}.

\subsubsection{Data Phase}
After the pilot block $\Xp$, the \APs transmit information-bearing symbols in the matrix $\Xd$, constructed as described in \Eqref{eq:stbc}. The terminal will, as in the case of perfect \CSI, detect the real and imaginary parts of the symbol separately based on the received signal vector \Eqref{eq:received:signal:vector} with the use of the channel estimate, $\hhat$, instead of the true channel, $\bh$.

For the real part, we have \cite{Karlsson18}
\begin{equation}\label{eq:detect:real}
\begin{aligned}
\hat{\bar{s}}_n
= \real{\hhat\herm\An\herm\by}
&= \real{\sqrt{\pd}\hhat\herm\An\herm\Xd\hhat
- \sqrt{\pd}\hhat\herm\An\herm\Xd\be
+ \hhat\herm\An\herm\bw}\\
&=\sqrt{\pd} \hhatnorm^{2}\rsn
+ \bar{\eta}_n
+ \bar{z}_n
\end{aligned}
\end{equation}
where
\[ \bar{\eta}_n \triangleq -\sqrt{\pd}\real{\hhat\herm\An\herm\Xd\be} \]
and
\[ \bar{z}_n \triangleq \real{\hhat\herm\An\herm\bw}. \]
The first error term in \Eqref{eq:detect:real}, $\bar{\eta}_n$, is due to the imperfect channel estimation (and is absent in \Secref{sec:snr:perfect:csi}). The second error term, $\bar{z}_n$, stems from the additive noise.
Similarly, for the imaginary part
\begin{equation}\label{eq:detect:imag}
\begin{aligned}
\hat{\tilde{s}}_n
&= \real{-\iu\hhat\herm\Bn\herm\by}
= \imag{\hhat\herm\Bn\herm\by}\\
&= \imag{\sqrt{\pd}\hhat\herm\Bn\herm\Xd\hhat
-  \sqrt{\pd}\hhat\herm\Bn\herm\Xd\be
+ \hhat\herm\Bn\herm\bw}\\
&=\sqrt{\pd} \hhatnorm^{2}\isn
+ \tilde{\eta}_n
+ \tilde{z}_n
\end{aligned}
\end{equation}
where
\[ \tilde{\eta}_n \triangleq -\sqrt{\pd}\imag{\hhat\herm\Bn\herm\Xd\be} \]
and
\[ \tilde{z}_n \triangleq \imag{\hhat\herm\Bn\herm\bw}. \]
From \Eqref{eq:detect:real} and \Eqref{eq:detect:imag}, we can write the complex, processed symbol as
\begin{equation}\label{eq:ls:processed:complex:symbol}
\hat{s}_n=\sqrt{\pd}\hhatnorm^{2}\sn +\eta_n + z_n,
\end{equation}
where $\eta_n = \bar{\eta}_n + \iu\tilde{\eta}_n$ and $z_n = \bar{z}_n + \iu\tilde{z}_n$.

At this stage, \Eqref{eq:ls:processed:complex:symbol} can be seen as the received symbol when the symbol $\sn$ is transmitted over the known channel $\sqrt{\pd}\hhatnorm^{2}$, contaminated by some additive noise. Since the noise term $\eta_n$ is correlated with the symbol $\sn$, we split it into a correlated and an uncorrelated part as
\[ \eta_n = c_n\sn + u_n, \]
where
\[ c_n \triangleq \dfrac{\Exp{\sn\conj\eta_n\given\hhat}}{\Es} \]
and
\[ u_n \triangleq \eta_n - c_n\sn. \]
The power of the uncorrelated noise $u_n$ is
\[ \Exp{|u_n|^{2}\given \hhat} = \Exp{|\eta_n|^{2}\given\hhat} - \Es|c_n|^{2}.\]
The received and processed symbol can then be written as
\begin{equation}\label{eq:ls:processed:complex:symbol:uncorrelated}
\hat{s}_n = \left(\sqrt{\pd}\hhatnorm^{2} + c_n\right)\sn + u_n + z_n, 
\end{equation}
which can be seen as the symbol $\sn$ passing through a known channel $\left(\sqrt{\pd}\hhatnorm^{2} + c_n\right)$, contaminated with uncorrelated noise $u_n + z_n$. The \SNR can be expressed as
\begin{equation}\label{eq:ls:snr}
\snr^{\LS}=\dfrac{\Es\left|\sqrt{\pd}\norm{\hhat}^{2} + c_n\right|^{2}}{\Exp{|\eta_n|^{2}\given\hhat} +  \Exp{|z_n|^{2}\given\hhat} - \Es|c_n|^{2}}.
\end{equation}

The following theorem gives the closed form expression of the \SNR, when using a general \ostbc:
\begin{theorem}\label{thm:ostbc:snr}
For a general \ostbc, we have
\begin{equation}\label{eq:cn:calculated}
c_n = -\sqrt{\pd}\left( \hhat\herm\Ucond\hhat + \iu\imag{\hhat\herm\An\herm\Bn\Ucond\hhat}\right),
\end{equation}
\[ \Exp{|z_n|^{2}\given\hhat} = \hhatnorm^{2}, \]
and
\begin{equation}\label{eq:noise:power:eta}
\Exp{|\eta_n|^{2}\given\hhat} = \dfrac{\pd\Es}{4}\left(\psi(\An,\bQ_1) + \bar{\psi}(\An,\bQ_2) + \psi(\Bn,\bQ_1) - \bar{\psi}(\Bn,\bQ_2)\right),
\end{equation}
where
\[ \psi(\bC,\bQ) \triangleq \real{\hhat\herm\bC\herm\left(\sum\limits_{k=1}^{\ns} \Ak\bQ\Ak\herm + \Bk\bQ\Bk\herm\right)\bC\hhat}, \]
\[ \bar{\psi}(\bC,\bQ) \triangleq \real{\hhat\herm\bC\herm\left(\sum\limits_{k=1}^{\ns} \Ak\bQ\Ak\trans - \Bk\bQ\Bk\trans\right)\bC^{*}\hhat^{*}}, \]
\[ \bQ_1 \triangleq \Exp{\be\be\herm\given\hhat} = \Ucond\hhat\hhat\herm\Ucond\herm + \Ccond, \]
and
\[ \bQ_2 \triangleq \Exp{\be\be\trans\given\hhat} = \Ucond\hhat\hhat\trans\Ucond\trans.\]
\begin{proof}
The proof is given in \Appref{app:ostbc:snr}.
\end{proof}
\end{theorem}

The insertion of \Eqref{eq:cn:calculated} and \Eqref{eq:noise:power:eta} into \Eqref{eq:ls:snr} yields a closed-form expression for the \SNR. However, this expression does not provide much intuition for how different parameters affect the \SNR. As a special case, when all \APs transmit the same message, we get a more palpable expression of the \SNR from the following corollary:

\begin{corollary}\label{cor:snr:one:group}
When $\Ng=1$ the \SNR in \Eqref{eq:ls:snr} reduces to
\begin{equation}\label{eq:snr:distribution:ls:siso}
\snr^{\LS}\sim \ExpDist{\lambda^{\LS}},
\end{equation}
where
\begin{equation}\label{eq:lambda:imperfect}
\lambda^{\LS}\triangleq \dfrac{1 + \bar{\beta}(\pp\tp+\pd\Es)}{\pd\Es\pp\tp\bar{\beta}^{2}}.
\end{equation}
\begin{proof}
The proof is given in \Appref{app:snr:one:group}.
\end{proof}
\end{corollary}

\changed{We can now explicitly see the effect of all relevant parameters, for example the pilot energy $\pp\tp$ and the data power $\pd$. By letting the pilot energy $\pp\tp$ grow large, the distribution of $\snr^{\LS}$ in \Eqref{eq:snr:distribution:ls:siso} approaches that of $\snr^{\textsc{p}}$ in \Eqref{eq:snr:distribution:perfect} (with one group), as is expected. In addition, from Corollary~\ref{cor:snr:one:group} we see that adding more \APs will always help, as $\prob{\snr^{\LS} < x}$  decreases with increasing $\bar{\beta}$ (cf. \Eqref{eq:beta:bar}) for a fixed $x$. That being said, adding an \AP far away from the terminal will have a negligible effect on the performance.}

Let us briefly study the difference in performance between having perfect \CSI and having to estimate the channel at the terminal. Assume that the transmit powers for pilots and data are identical ($\pp=\pd=\rho$ in \Eqref{eq:lambda:imperfect}) and that all symbols have unit energy ($\Es=1$). Then 
\begin{equation}\label{eq:lambda:ls:example}
	\lambda^{\LS}= \dfrac{1 + \rho\bar{\beta}(1+\tp)}{\rho^{2}\tp\bar{\beta}^{2}}.
\end{equation}
Comparing \Eqref{eq:lambda:ls:example} to \Eqref{eq:lambda:perfect}, we see that the performance with perfect \CSI at the terminal and a transmit power of $\rho$ is identical to the performance with imperfect \CSI at the terminal and a transmit power of 
\[\rho\left(\dfrac{1+\tp}{2\tp} + \sqrt{\dfrac{(1+\tp)^{2}}{4\tp^{2}} + \dfrac{1}{\rho\bar{\beta}\tp}}\right).  \] 
When $\rho\bar{\beta}$ is large, perfect \CSI has an advantage of approximately
\[\dfrac{1+\tp}{\tp}, \]
which is about $3$~dB if a single pilot symbol is used.

\section{Performance Metric}\label{sec:performance:metric}

Instead of ergodic rates, we will measure performance in terms of outage rates. A link is said to be in outage if it does not satisfy the \SNR constraint $\snr \geq \gamma$ for some threshold $\gamma>0$. When this happens, the rate at which the transmitter is sending data is larger than what the channel supports. The probability of this occurring is the outage probability, defined as
\[ p^{\text{out}}(\gamma) \triangleq \prob{\snr < \gamma}. \]

We define the outage rate as
\begin{equation}\label{eq:performance:metric:outage:rate}
R_{\epsilon}\triangleq \left(1-\dfrac{\tp}{\tc}\right)\dfrac{\Ns}{\td}\log_2\left(1+ \gamma_{\epsilon}\right)\ \bpcu,
\end{equation}
where $\gamma_{\epsilon}$ satisfies
\[ p^{\text{out}}(\gamma_{\epsilon}) = \epsilon. \]
The first factor in \Eqref{eq:performance:metric:outage:rate} is the fraction of the coherence interval, consisting of $\tc$ samples, spent on data and the second factor is the \emph{code rate} of the \ostbc used, as $\Ns$ symbols are transmitted over $\td$ channel uses.

In order to calculate the outage rate for a specific outage probability, $\epsilon$, we need to find the value of $\gamma_{\epsilon}$, which depends on the distribution of the \SNR. From \Secref{sec:snr}, we have closed-form expressions for the \SNR. For some special cases we further have the distribution of the \SNR in closed form (for fixed large-scale fading): \Eqref{eq:snr:distribution:perfect} for the case of perfect \CSI at the terminal and \Eqref{eq:snr:distribution:ls:siso} for the case when all \APs transmit the same symbol ($\Ng=1$) and the terminal estimates the channel using \LS.

Recall that the closed-form expressions for the \SNR distributions depend on the large-scale fading; thus, in order to find the \SNR threshold for a random terminal in the network, we need to incorporate the randomness of this large-scale fading. When the terminal has perfect \CSI, the coverage probability is given by
\begin{equation}\label{eq:coverage:probability:perfect:csi}
\prob{\snr^{\textsc{p}}\geq \gamma} = \Exp{\prob{\snr^{\textsc{p}}\geq \gamma\given \{\bar{\beta}_n\}}} = \Exp{\sum\limits_{n=1}^{\Ng}\dfrac{e^{-\gamma\lambda^{\textsc{p}}_n}}{\prod\limits_{k\neq n}^{\Ng}\left(1 - \dfrac{\lambda^{\textsc{p}}_n}{\lambda^{\textsc{p}}_k}\right)}}, 
\end{equation}
where the expectation is over the randomness of the large-scale fading. We have used the fact that the \SNR is distributed as a sum of exponential distributions (hyperexponential), whose probability density function is known, see for example \cite{Bjornson09}. In \Eqref{eq:coverage:probability:perfect:csi}, it is assumed that $\bar{\beta}_n\neq\bar{\beta}_m$ if $n\neq m$, which happens with probability one in practice.

With an estimated channel at the terminal and a single \AP group ($\Ng=1$), the coverage probability is given by
\begin{equation}\label{eq:coverage:probability:ls:single:group}
\prob{\snr^{\LS}\geq \gamma} = \Exp{e^{-\gamma\lambda^{\LS}}}. 
\end{equation}

For both \Eqref{eq:coverage:probability:perfect:csi} and \Eqref{eq:coverage:probability:ls:single:group}, the coverage probability depends only on the distribution of the large-scale coefficients, and can be calculated relatively quickly through Monte-Carlo simulations. For the remaining cases of interest, when the terminal estimates the channel and we have more than one \AP group, the \SNR in \Eqref{eq:ls:snr} has to be Monte-Carlo simulated, which includes simulating the small-scale fading. From a computational perspective, \Eqref{eq:coverage:probability:perfect:csi} and \Eqref{eq:coverage:probability:ls:single:group} should be used whenever possible, as simulating \Eqref{eq:ls:snr} requires considerably more time.

\section{Numerical Evaluation}\label{sec:numerical:evaluation}

In this section, we will evaluate the coverage performance of a cell-free massive \MIMO system, in terms of outage rates and received \SNR. There are many different aspects to consider, so in order to make the different comparisons as clear as possible, we focus on one aspect at a time. We will start with a basic scenario, using a simple model. We will later add new features to this model and discuss each feature separately.

The basic scenario is defined as follows: The terminals are assumed to have perfect \CSI, there is no shadow fading, and no transmit diversity is used (all \APs are in the same group). In the following sections, we will analyze the effects of \AP distribution, \AP density, shadow fading, channel estimation, \AP grouping, pilot/data power optimization, transmit diversity, receive diversity, and \APs with multiple antennas.

The parameters that will be fixed throughout all simulations are as follows: The coherence bandwidth is $B_{\textsc{c}} = 200$ kHz and the coherence time is $1.5$ ms, resulting in a coherence interval of $\tc=300$ samples. The transmit power per \AP is $p = 1$~mW over the bandwidth $B_{\textsc{c}}$, corresponding to a transmit power of $100$~mW over $20$ MHz bandwidth. The normalized energy budget for one coherence interval is $E\triangleq\rho\tc$, where $\rho$ is the normalized transmit power defined as
\[ \rho \triangleq  p / (B_{\textsc{c}} T k_{\text{B}}F), \]
where $T = 300$ K is the surrounding temperature, $k_{\text{B}}$ is Boltzmann's constant, and $F= 9$ dB is the noise figure. The outage probability is set to $\epsilon = 10^{-3}$.

We adopt a three-slope variant of the COST Hata model for the path loss \cite{Rappaport01}, as in, for example \cite{Ngo17a}. In particular this means that the path loss (measured in dB) at distance $d$ km from an \AP is
\[ \begin{cases}
L + 15\log_{10}(d_{\textsc{o}}/d_{\textsc{r}}) + 20\log_{10}(d_{\textsc{i}}/d_{\textsc{r}}),& \text{ if } d\leq d_{\textsc{i}}\\
L + 15\log_{10}(d_{\textsc{o}}/d_{\textsc{r}}) + 20\log_{10}(d/d_{\textsc{r}}),& \text{ if } d_{\textsc{i}}<d\leq d_{\textsc{o}}\\
L + 35\log_{10}(d/d_{\textsc{r}}),& \text{ if } d> d_{\textsc{o}},
\end{cases} \]
where
\newcommand{\fc}{f_{\textsc{c}}}
\newcommand{\hAP}{h_{\AP}}
\newcommand{\hT}{h_{\textsc{t}}}
\[ L\triangleq 46.3 
	+ 33.9\log_{10}(\fc) 
	- 13.82\log_{10}(\hAP) 
	- (1.1\log_{10}(\fc) - 0.7)\hT
	+ 1.56\log_{10}(\fc)-0.8.\]
$L$ represents the path loss in dB experienced at some reference distance $d_{\textsc{r}}$ km. Moreover, $\fc$ is the carrier frequency in MHz, $\hAP$ is the height of an \AP in meter, and $\hT$ is the height of the terminal in meter. Here, we choose $\fc = 1.9$~GHz, $\hAP=15$~m, $\hT=1.5$~m. This gives a path loss of about $L=141$~dB at $d_{\textsc{r}} = 1$~km.
	
The distances $d_{\textsc{i}}$ and $d_{\textsc{o}}$ determine where the ``slope'', or the path-loss exponent, changes. The effective path-loss exponent increases in steps with the distance, until $d=d_{\textsc{o}}$. For $d>d_{\textsc{o}}$, the effective path-loss exponent is 3.5. In the simulations, we have $d_{\textsc{i}} = 10$~m and $d_{\textsc{o}}=50$~m. 

\changed{Note that in an infinitely large network with randomly distributed \APs and terminals, all terminals are statistically identical if the two distributions are mutually independent. This allows us to consider the performance of a single user terminal, located at the origin, surrounded by a large number of geographically distributed \APs. The considered network is large enough network to eliminate any edge effects and the results in the numerical examples below are obtained by considering many independent realizations of the network.}

\subsection{Distribution of Access Points}\label{sec:ap:distribution}

We consider two different ways of placing \APs over a specific area. The first way is the classical hexagonal grid. Here, all \APs have the same distance to all of its six neighbors and the lattice of \APs follows a predefined pattern. We expect the hexagonal lattice to perform well, as the pattern minimizes the maximum distance from an arbitrary terminal, to the closest \AP for a given \AP density (\APs per km$^2$). This should be ideal from a coverage-probability perspective.

Second, we consider \AP locations drawn from a two-dimensional Poisson-point process (\PPP). This way of distributing base stations in an area was popularized in \cite{Andrews11}. The argument for this distribution is that it more closely resembles the topology of a practical system, where the \APs cannot be placed on a hexagonal grid. It also matches well with real-world deployments \cite{Lu15}. Because of the randomness of the \AP locations, large holes might appear, where it is far from a terminal to the closest \AP. For this reason, as we are concerned with low percentiles, the performance is expected to be worse than for the hexagonal lattice.

We do not add any repulsive constraints on the \PPP, so the \APs could be arbitrarily close to each other. However, as argued in \cite{ElSawy17}, the absolute positions of the \APs do not matter, only their relative position to the terminal, which is assumed to be in the origin. Moreover, the path loss model makes sure that the model does not break down, even if the terminal is very close to the nearest \AP, since the path loss is constant distances shorter than $d_{\textsc{i}}$. In addition, it is the least favored terminals, located far away from the \APs that will limit the coverage performance; thus terminals in close proximity to the \APs will have little influence on the overall performance.

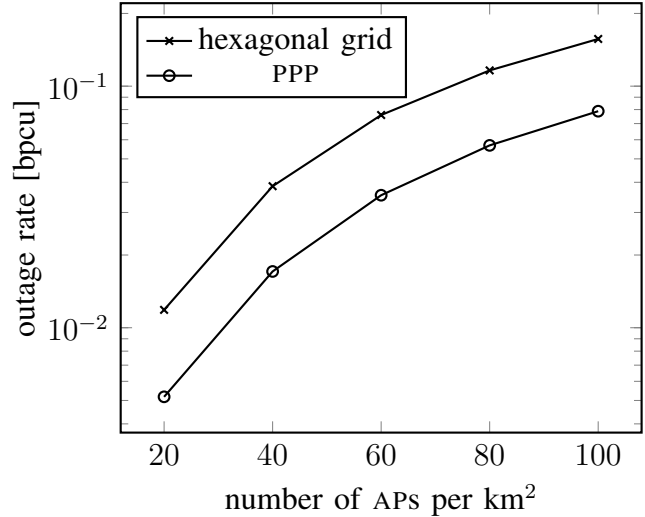
\begin{figure}[ht]
\centering
\begin{tikzpicture}
	\begin{axis}[
	xlabel={number of \APs per km$^{2}$},
	ylabel={outage rate [bpcu]},
	legend pos = {north west},
	table/col sep=comma,
	ymode=log,]
	\addplot[black,thick, solid, mark=x] table[x index={0},y index={1}] {figures/data-for-paper/hex-vs-ppp.dat};
	\addlegendentry{hexagonal grid};
	\addplot[black,thick, solid, mark=o] table[x index={0},y index={2}] {figures/data-for-paper/hex-vs-ppp.dat};
	\addlegendentry{\PPP};
	\end{axis}
\end{tikzpicture}
	\caption{The outage rate for two different ways of distributing \APs over an area: one random, according to a \PPP, and one in a fixed, hexagonal lattice. The hexagonal lattice is better than the random deployment because it minimizes the largest distance between an arbitrary terminal and the closest \AP for a fixed \AP density. For the \PPP, large holes between \APs might appear, causing a drop in the outage rate compared to the hexagonal lattice.}\label{fig:hex:vs:ppp}
\end{figure}

A comparison of the two distributions is shown in \Figref{fig:hex:vs:ppp}, for different \AP densities, i.e., number of \APs per km$^2$. We see, as expected, that distributing the \APs evenly over the area, in a hexagonal fashion leads to larger outage rates, for any density. In the following, we restrict ourselves to \APs distributed according to a \PPP, as this models real wold deployment fairly accurately \cite{Andrews11,Lu15} while simultaneously giving a lower bound on performance. 

\changed{It is no surprise that the outage rates increase with the \AP density, as adding more \APs in an area always increases the probability of coverage since there is no interference. In the following, we will assume a density of 20 \APs per km$^{2}$, corresponding to distance of about 240 meters between neighboring \APs on a hexagonal lattice. All results given below would be improved if the \AP density were higher.}

\subsection{Modeling the Large-scale Fading}

There are many ways to model the large-scale fading, from the simple models with only distant-dependent path loss, to more sophisticated models based on measurements. The simple models might not model reality as well as the more sophisticated ones, but can still be used to get some insight. When dealing with random \AP placements, using the path-loss-only model may give very neat expressions in closed form, which in turn give insights, such as how frequency reuse affect the coverage of active terminals \cite{Andrews11}. We will now analyze how the coverage performance for inactive terminals in a cell-free massive \MIMO setup changes depending on the large-scale fading model. 

We consider the simplest model with only path loss and two models including shadow fading: one uncorrelated and one correlated. We let $v_{k,m}$ denote the loss in dB due to shadowing in the link between terminal $k$ and \AP $m$. For the uncorrelated shadow fading, $v_{k,m}~\sim~\normal{0}{\stdshadow^{2}}$, with $v_{k,m}$ and $v_{k',m'}$ mutually independent for $(k,m)\neq(k',m')$. If the shadow fading is correlated, we let
\[ v_{k,m} = \sqrt{\delta}a_k + \sqrt{1 - \delta}b_m, \]
where $a_k$ and $b_m$ are $\normal{0}{\stdshadow^{2}}$ and uncorrelated, as in \cite{Ngo17a}. Moreover, $0\leq\delta\leq 1$. The two shadow fading coefficients have the following correlation:
\[ \Exp{a_k a_{k'}} = 2^{-\frac{d_a(k,k')}{d_{\textsc{u}}}} \text{ and } 
\Exp{b_m b_{m'}} = 2^{-\frac{d_b(m,m')}{d_{\textsc{u}}}},\]
where $d_a(k,k')$ is the distance between terminals $k$ and $k'$, and $d_b(m,m')$ is the distance between \APs $m$ and $m'$. The constant $d_{\textsc{u}}$ is termed the decorrelation distance and specifies the spatial correlation. For a given distance, two shadow-fading coefficients are less correlated the smaller the decorrelation distance is. The uncorrelated case can be seen as a special case, letting $d_{\textsc{u}}$ tend to zero. In the simulations, we set $\delta = 0.5$ so that only the spatial separation matters, regardless if it is between two \APs or two terminals. Moreover, we set $d_{\textsc{u}}=0.2$~km and $\stdshadow=8$, which are typical values for suburban environments \cite{Gudmundson91}.

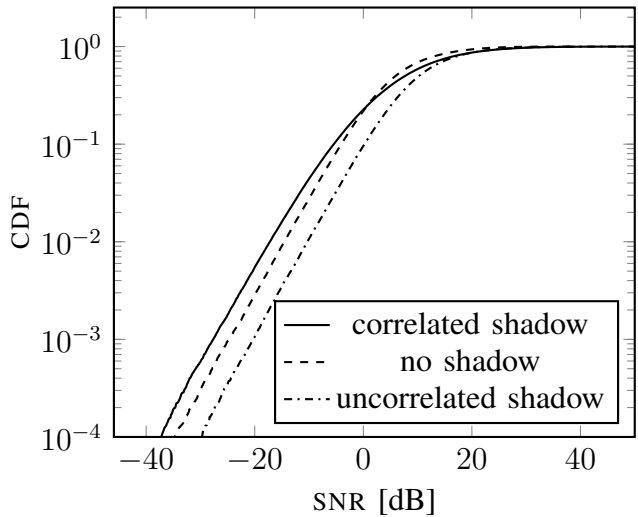
\begin{figure}[ht]
\centering
\begin{tikzpicture}
	\begin{axis}[
		xlabel={\SNR [dB]},
		ylabel=\CDF,
		ymode=log,
		ymin=1e-4,
		xmax=50,
		legend pos = {south east},
		table/col sep=comma,
		table/x index=1,
		table/y index=0,
		]
	\addplot[thick, solid] table {figures/data-for-paper/snrs-correlated.dat};
	\addlegendentry{correlated shadow};
	\addplot[thick, dashed] table {figures/data-for-paper/snrs-path-loss.dat};
	\addlegendentry{no shadow};
	\addplot[thick, dashdotted] table {figures/data-for-paper/snrs-shadow.dat};
	\addlegendentry{uncorrelated shadow};
	\end{axis}
\end{tikzpicture}
	\caption{The \SNR of an arbitrary terminal in a cell-free massive \MIMO system for different large-scale-fading models: no shadow fading, uncorrelated shadow fading, and correlated shadow fading. For low percentiles, the model including correlated shadow fading, which is arguably the most accurate, gives lower performance than the other models. The performance is improved with uncorrelated shadow fading over the case with only path loss because of the large amount of macro diversity.}\label{fig:large:scale:fading}
\end{figure}

The comparison of the three different models is shown in \Figref{fig:large:scale:fading}. Since we are interested in outage rates and coverage probability, it is the lower tail of the \CDF that is of concern here, in particular the \SNR value corresponding to $\epsilon=10^{-3}$.

At first glance, the results may seem unintuitive: adding shadow fading improves performance. But recall that the median of the shadow fading is unity, which means that half of the large-scale coefficients will increase when adding shadow fading. Because of the abundance of macro diversity, it is likely that at least one of the large-scale coefficients for \APs close to a certain terminal have increased, thus the \SNR improves. When correlating the shadow fading, the performance drops, because the macro diversity is not as prevalent. It is now more likely that all \APs close to a terminal all have a decreased large-scale coefficient to that terminal. In the following, the large-scale fading will include correlated shadow fading.

\subsection{Estimating the Channel}\label{sec:estimating:the:channel}

\begin{figure}[ht]
	\centering
	\begin{tikzpicture}
	\begin{axis}[xlabel={number of pilots, $\tp$},
	ylabel={outage rate [bpcu]},
	table/col sep=comma,
	legend pos={south west},
	]
	\addplot[black, solid] table[x index={0},y index={3}] {figures/data-for-paper/csi-vs-nocsi.dat};
	\addlegendentry{perfect \CSI};
	\addplot[black, dashed] table[x index={0},y index={2}] {figures/data-for-paper/csi-vs-nocsi.dat};
	\addlegendentry{$\tp=1$, opt. $\pp$};
	\addplot[black, dashdotted] table[x index={0},y index={1}] {figures/data-for-paper/csi-vs-nocsi.dat};
	\addlegendentry{$\pp=\pd$};
	\end{axis}
\end{tikzpicture}
	\caption{The outage rates when the terminal knows the channel (has perfect \CSI) compared to when the terminal estimates the channel. The dash-dotted line shows the outage rate when varying the number of pilot symbols while keeping the pilot power and data power equal ($\pd=\pp$). The outage rate increases at first, as the increased number of pilots increases the accuracy of the channel estimate. After a certain point adding more pilots do not increase the rate enough to justify the increased pilot overhead, and the outage rate decreases. The dashed line shows the outage rate when the pilot/data power is optimized, while using a single pilot symbol. }\label{fig:csi:vs:nocsi}
\end{figure}
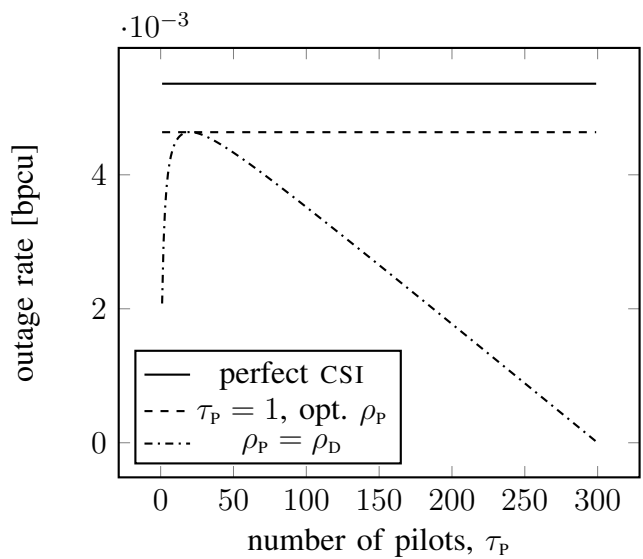

In a practical system, the terminals will not have perfect \CSI and will have to estimate the channel. Apart from not knowing the channel perfectly, the added pilot overhead means less room for data. The effect of channel estimation, for different pilot lengths is shown in \Figref{fig:csi:vs:nocsi}. We see that increasing the number of pilots is beneficial at first, as the increase in energy spent on pilots $\pp\tp$ increases the \SNR. After a certain point, however, the benefits of increasing the energy spent on pilots is overcome by the penalty of the increased pilot overhead. 

To increase the energy spent on pilots, $\pp\tp$, without increasing the pilot overhead we consider optimizing the  distribution of power between the pilot power, $\pp$, and the data power, $\pd$. Since each \AP has a fixed energy budget, $E$, to spend in one coherence interval, increasing the pilot power will decrease the data power. We have the following relation:
\begin{equation}\label{eq:power:relationship}
\pd = (E - \pp\tp)/(\tc-\tp).
\end{equation}
It is challenging to do this optimization since the \APs do not have any \CSI, instantaneous or statistical. We employ the heuristic method described below. This method follows the same general strategy as the power optimization described in \cite{Karlsson18}.

First, assume that all \APs are in the same group. From this we know that the \SNR distribution is given by \Eqref{eq:snr:distribution:ls:siso}. Thus, maximizing the probability of coverage is equivalent to minimizing $\lambda^{\LS}$ in \Eqref{eq:lambda:imperfect}. This is done by inserting \Eqref{eq:power:relationship} into \Eqref{eq:lambda:imperfect}, differentiating with respect to $\pp$, and equating the result to zero. Thus, optimal power distribution would be straightforward, if all terminals had the same large-scale fading $\bar{\beta}$, and if this large-scale fading were known to the \APs; unfortunately, neither is true.

In order to distribute the power between data and pilots, even without knowledge of the large-scale fading, the following method is employed:
\begin{enumerate}
\item Find the position within the network that is furthest away from the closest \AP (largest minimum distance to any of the \APs).\footnote{This itself is a non-trivial problem. In the implementation, a grid search is used to find a ``bad'' position, but not necessarily the worst.} Call this position $t_{\textsc{w}}$ (for ``worst'').
\item Calculate the large-scale coefficient associated with $t_{\textsc{w}}$, denoted $\beta_{\textsc{w}}$, only assuming distance-dependent path loss (i.e., no shadow fading).
\item Find the optimal power distribution for this large-scale coefficient by inserting $\beta_{\textsc{w}}$ into \Eqref{eq:lambda:imperfect}, the expression for $\lambda^{\LS}$, and minimize this as described above.
\end{enumerate}

Note that the assumptions made in this heuristic method ($\Ng=1$ and no shadow fading) do not need to hold for the method to be useful. Moreover, the only information needed to perform this optimization is the locations of the \APs.

The outage rate when optimizing the pilot power is also shown in \Figref{fig:csi:vs:nocsi}, when the minimum number of pilots ($\tp=1$ in this case) is used. In \Figref{fig:csi:vs:nocsi} we see that choosing the number of pilots giving the highest outage rate while keeping the data and pilot power equal gives the same result as performing our heuristic optimization on the pilot power. Thus, in this particular case, optimizing the power or the number of pilots would not make any difference when looking at outage rates. However, the difference between the two optimization methods grows with \SNR. In general, optimizing the power gives better performance than optimizing the number of pilots \cite{Cheng17}. However, because of the heuristic nature of the optimization, optimizing the power $\pp$ is not necessarily better than doing an exhaustive search over the number of pilots, $\tp$. In the following, we assume the minimum number of pilots is used, $\tp = \Ng$, together with the power optimization described in this section, unless explicitly stated otherwise. 

As a result of the power optimization, the pilot power will be considerably higher than the data power. However, if the pilot symbols are adequately spread out in time and frequency, the peak-to-average power ratio will still be tolerable.

\subsection{Transmit Diversity}\label{sec:transmit:diversity}

\changed{To increase the reliability and the outage rate, we now let the \APs jointly transmit an \ostbc. We consider two square codes, namely the Alamouti code \cite{Alamouti98} and a four dimensional code with rate $3/4$ \cite{Larsson03,Karlsson18}.}

\changed{Before transmission can occur, each \AP needs to know what group it belongs to, in order to know what part of the \ostbc to transmit. The simplest way to group \APs is to randomize the grouping: randomly assign $1/\Ng$ of the \APs to each of the $\Ng$ groups. Technically, there is no constraint that each group must contain the same number of \APs, but it is intuitively reasonable to not favor one part of the \ostbc over the other. This grouping problem does not appear in conventional massive \MIMO, since the \AP (base station) has enough antennas to be able to transmit the entire \ostbc by itself.}

\begin{figure}[ht]
	\centering
	\begin{tikzpicture}
	\begin{axis}[xlabel={rate [bpcu]},
	ylabel={\CDF},
	table/col sep=comma,
	legend pos={north west},
	ymode=log,
	xmode=log,
	ymin=5e-4,
	ymax=5e-2,
	]
	\addplot[black, thick, solid, smooth] table[x index={7}, y index={3}] {figures/data-for-paper/transmit-diversity.dat};
	\addlegendentry{nominal};
	\addplot[black, thick, dashed, smooth] table[x index={6}, y index={2}] {figures/data-for-paper/transmit-diversity.dat};
	\addlegendentry{$\Ng=1$};
	\addplot[black, thick, dashdotted, smooth] table[x index={5}, y index={1}] {figures/data-for-paper/transmit-diversity.dat};
	\addlegendentry{$\Ng=2$};
	\addplot[black, thick, dotted, smooth] table[x index={4}, y index={0}] {figures/data-for-paper/transmit-diversity.dat};
	\addlegendentry{$\Ng=4$};
	\end{axis}
\end{tikzpicture}
	\caption{\changed{The rate for an arbitrary terminal is increased by optimizing the pilot power and adding spatial diversity with an \ostbc. For the chosen operating point, $\epsilon=10^{-3}$, optimizing the pilot power brings a gain corresponding to about $6$ dB, and using transmit diversity brings an additional $6$ dB and $8$ dB gain for $\Ng=2$ and $\Ng=4$, respectively.}}\label{fig:transmit:diversity}
\end{figure}
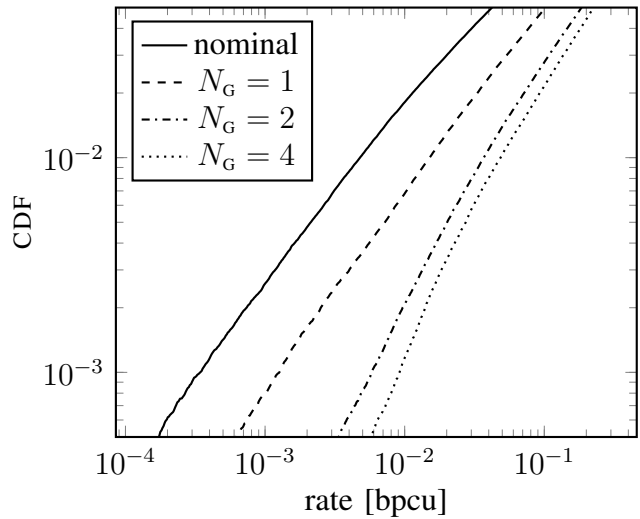

\changed{The benefits of adding transmit diversity in form of an \ostbc are illustrated in \Figref{fig:transmit:diversity}, where the \CDF of the rate for an arbitrary terminal in the system is shown when random grouping is used. As a reference, the nominal case of a single group and no power optimization is shown. At the $10^{-3}$ percentile, transmitting with the Alamouti code ($\Ng=2$) and the larger code ($\Ng=4$) bring gains corresponding to approximately $12$~dB and $14$~dB, respectively, over the nominal case with $\Ng=1$ and no power optimization. The gain from using a larger code increases with lower percentiles.}

\subsection{Grouping the Access Points}\label{sec:other:colorings}

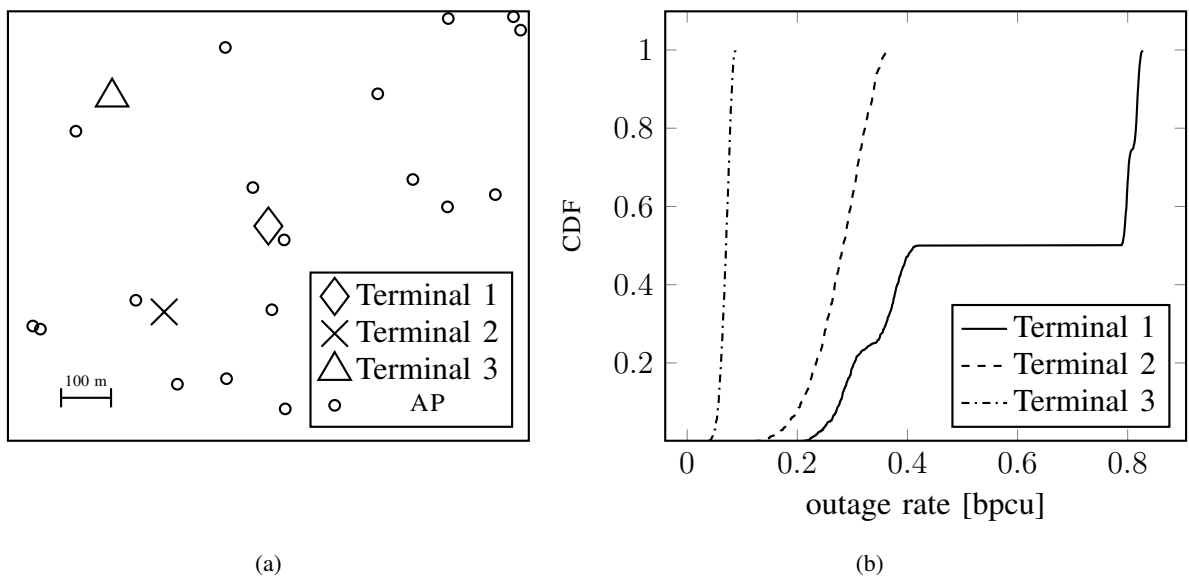
\begin{figure}[ht]
\centering
	\subfloat[]{\label{fig:random:coloring:positions}
	\begin{tikzpicture}
		\def\markSize{7pt}
		\begin{axis}[
		xtick={\empty},
		ytick={\empty},
		table/col sep=comma,
		table/x index=0,
		table/y index=1,
		xmin=-0.5,
		xmax=0.5,
		ymin=-0.5,
		ymax=0.5,
		only marks,
		xlabel=\phantom{outage rate [bpcu]},
		legend pos={south east},]
		\addplot[mark=diamond, mark size=\markSize] table {figures/data-for-paper/ue-pos0.dat};
		\addlegendentry{Terminal 1};
		\addplot[mark=x, mark size=\markSize] table {figures/data-for-paper/ue-pos2.dat};
		\addlegendentry{Terminal 2};
		\addplot[mark=triangle, mark size=\markSize] table {figures/data-for-paper/ue-pos1.dat};
		\addlegendentry{Terminal 3};
		\addplot[mark=o] table {figures/data-for-paper/ap-pos.dat};
		\addlegendentry{\AP};
		\coordinate (left) at (axis cs:-0.4,-0.4);
		\coordinate (right) at (axis cs:-0.3,-0.4);
		\draw[|-|] (left) -- node[midway,above] {\tiny{100 m}} (right);
		\end{axis}
\end{tikzpicture}}
	\subfloat[]{\label{fig:random:coloring:cdfs}
	\begin{tikzpicture}
		\begin{axis}[xlabel={outage rate [bpcu]},
		ylabel={\CDF},
		table/col sep=comma,
		table/x index=0,
		table/y index=1,
		legend pos={south east},
		ymin=1e-3,
		]
		\addplot[black, solid, thick] table {figures/data-for-paper/random-coloring0.dat};
		\addlegendentry{Terminal 1};
		\addplot[black, dashed, thick] table {figures/data-for-paper/random-coloring2.dat};
		\addlegendentry{Terminal 2};
		\addplot[black, dashdotted, thick] table {figures/data-for-paper/random-coloring1.dat};
		\addlegendentry{Terminal 3};
		\end{axis}
\end{tikzpicture}}
		\caption{The effect of grouping the \APs randomly for different terminals. Here we show a system with three terminals. The grouping can significantly affect the outage rate for a single terminal and usually has the largest impact on terminals close to one or a few \APs. The jump at the median for Terminal 1 in \Figref{fig:random:coloring:cdfs} occurs because two \APs are considerably closer to this terminal than other \APs, as seen in \Figref{fig:random:coloring:positions}. The outage rate is improved if these two \APs are in different groups. For terminals far away from the closest \AP, only a marginal increase in outage rate is observed.}\label{fig:random:coloring}
\end{figure}

\changed{To analyze the performance of the random grouping used in \Secref{sec:transmit:diversity} we consider an example scenario with a number of \APs transmitting the Alamouti code (${\Ng=2}$) to three terminals, depicted in \Figref{fig:random:coloring:positions}. The positions of the \APs and the terminals are fixed throughout this example. To not convolute the analysis, we temporarily give the terminals perfect \CSI and neglect the effect of shadow fading; thus, the only randomness in the outage rates shown in \Figref{fig:random:coloring:cdfs} is due to the grouping. For terminals far away from the nearest \AP, like Terminal $3$ in \Figref{fig:random:coloring}, there is a higher probability that there are more \APs at similar distances; hence, the specific grouping makes little difference. For terminals close to several \APs, like Terminal $1$ and $2$ in \Figref{fig:random:coloring} the \SNR is higher, leading to a larger outage rate. When a terminal is very close to more than one \AP, it is important that not all these \APs are in the same group. This can be seen by the discontinuity in the \CDF for Terminal $1$. In this case, the highest outage rates are achieved when the two closest \APs are in different groups.}

\changed{Is it possible to improve the performance by grouping the \APs in a more sophisticated way? Looking at Terminal 1 in \Figref{fig:random:coloring}, it looks plausible. Based on \Figref{fig:random:coloring} and other numerical experiments we conducted, grouping the \APs to maximize the smallest of the large-scale fading coefficients seems to be an effective strategy. With this in mind, we suggest a heuristic alternative to the random grouping with the aim of assigning \APs close to each other to different groups. We call this strategy \emph{neighbor grouping}, and it works as follows:
\begin{enumerate}
\item Calculate the pairwise distance between all \APs.
\item Let $n_{\textsc{a}}$ denote the number of \APs assigned to a group.
\item Locate the \AP pair with the minimum distance.
\item Assign one of these \APs to group $0$, set $n_{\textsc{a}} = 1$.
\item Denote the previously assigned \AP by $n_{\textsc{p}}$.
\item From $n_{\textsc{p}}$, locate the closest \AP that does not belong to any group, place it in group ${n_{\textsc{a}} \mod \Ng}$, and increment $n_{\textsc{a}}$ by $1$.
\item End if all \APs are in a group, else go to step 5.
\end{enumerate}}

\begin{figure}[ht]
	\centering
	\begin{tikzpicture}[spy using outlines={rectangle, magnification=3.3, size=2cm, connect spies}]
	\begin{axis}[
	xlabel={rate [bpcu]},
	ylabel={\CDF},
	table/col sep=comma,
	legend pos={north west},
	ymode=log,
	xmode=log,
	ymin=5e-4,
	ymax=5e-2,
	]
		\addplot[black, thick, solid, smooth] table[x index={5}, y index={1}] {figures/data-for-paper/smart-coloring.dat};
		\addlegendentry{$\Ng=2$, random};
		
		\addplot[black, thick, dashed, smooth] table[x index={4}, y index={0}] {figures/data-for-paper/smart-coloring.dat};
		\addlegendentry{$\Ng=4$, random};
		
		\addplot[black, thick, dotted, smooth] table[x index={7}, y index={3}] {figures/data-for-paper/smart-coloring.dat};
		\addlegendentry{$\Ng=2$, neighbor}
		
		\addplot[black, thick, dashdotted, smooth] table[x index={6}, y index={2}] {figures/data-for-paper/smart-coloring.dat};
		\addlegendentry{$\Ng=4$, neighbor};
	\begin{scope}
		\spy[size=2.5cm] on (2.6,1.9) in node at (5.2,1.6);
	\end{scope}
	\end{axis}
\end{tikzpicture}
	\caption{\changed{The rate for an arbitrary terminal when using random and neighbor grouping for two different \ostbcs. We see that random grouping with two groups performs slighly worse than the three other transmission strategies. The effect of the grouping is not as prominent as one might think, as the grouping makes little difference for terminals in unfavorable positions in general (cf. \Figref{fig:random:coloring}}).}\label{fig:smart:coloring:snrs}
\end{figure}
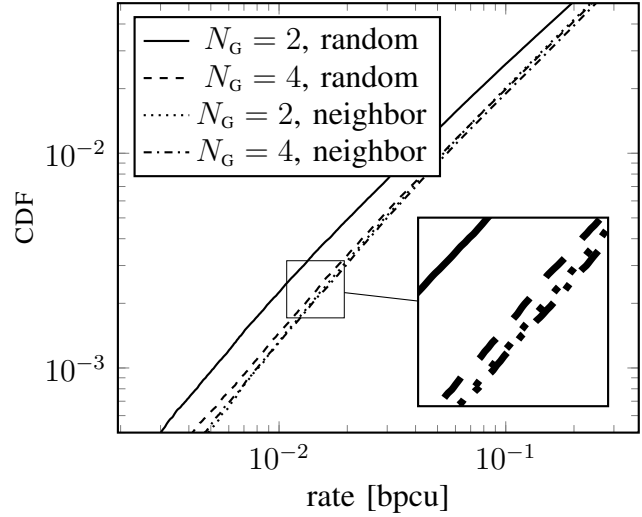

\changed{Neighbor grouping is compared to random grouping in \Figref{fig:smart:coloring:snrs} for the two considered \ostbcs. In this case, the neighbor grouping brings some gains for the smaller code, but no noticeable gains for the larger code. The corresponding difference between the worst ($\Ng = 2$, random grouping) and the best ($\Ng=4$, neighbor grouping) at the operating point $\epsilon~=~10^{-3}$ is about $2$ dB. Thus, according to our analysis, randomly assigning the \APs to a group works rather well, when the \APs are distributed according to a \PPP, especially for larger codes. When considering power optimization, choosing a code, and grouping the \APs, the former two seem to have a more prominent effect on the coverage performance than the latter.}

\changed{Some intuition to why grouping the \APs does not bring a larger gain can be found by looking back at \Figref{fig:random:coloring:cdfs}. The system's outage/coverage performance depends on the performance of the terminals in the worst positions. For a terminal in a bad position, such as Terminal 3 in \Figref{fig:random:coloring:positions}, the grouping makes little difference. That is, whether Terminal 3 operates at the $5$ or $95$ percentile in \Figref{fig:random:coloring:cdfs}, may not change the outage performance of the system noticeably. The larger the code is, the larger (on average) is the distance from the terminal to the group furthest away, implying that the grouping matters less.}

\subsection{Receive Diversity}

\changed{Let us digress for a moment. Up until now, we have focused on simple receivers with a single antenna because those terminals will be the bottle neck of the system in terms of coverage. However, many devices today are equipped with multiple antennas, which may be used for receive diversity. In this subsection, we briefly analyze the effect of multiple antennas at the terminal.}

\changed{We assume the same transmit strategy as before, but the receiver now has two antennas. It is assumed that the two channels $\bh_1$ and $\bh_2$ associated with the two antennas are uncorrelated, and that the processing of the received signal on each of the antennas is done separately. The processed signals for the two antennas are given by
\[ \hat{s}_{n,1} = \left(\sqrt{\pd}\norm{\hat{\bh}_1}^{2} + c_{n,1} \right)\sn + u_{n,1} + z_{n,1}  \]
and
\[ \hat{s}_{n,2} = \left(\sqrt{\pd}\norm{\hat{\bh}_2}^{2} + c_{n,2} \right)\sn + u_{n,2} + z_{n,2},  \]
respectively, where all variables are defined analogously to the ones in \Eqref{eq:ls:processed:complex:symbol:uncorrelated}. The terminal now performs maximum-ratio combining of the two processed symbols:
\[ \hat{s}_n = \left(\sqrt{\pd}\norm{\hat{\bh}_1}^{2} + c_{n,1} \right)^{*}\hat{s}_{n,1} + \left(\sqrt{\pd}\norm{\hat{\bh}_2}^{2} + c_{n,2} \right)^{*}\hat{s}_{n,2}  \]
which gives receive diversity, in addition to the transmit diversity obtained from the \ostbc.}

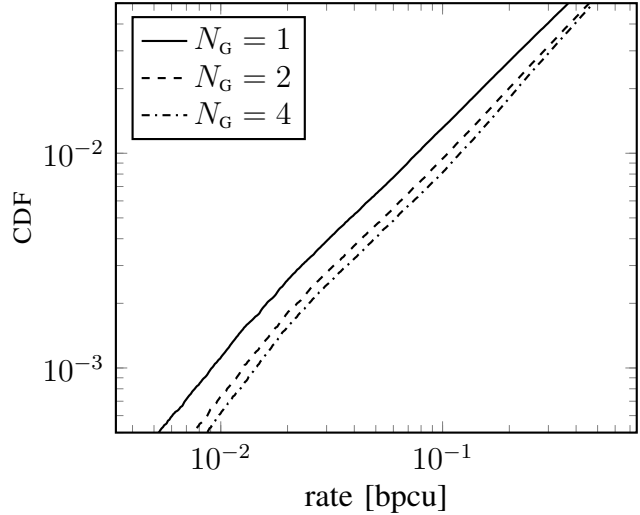
\begin{figure}[ht]
	\centering
	\begin{tikzpicture}
	\begin{axis}[xlabel={rate [bpcu]},
	ylabel={\CDF},
	table/col sep=comma,
	legend pos={north west},
	ymode=log,
	xmode=log,
	ymin=5e-4,
	ymax=5e-2,
		]
	
	\addplot[black, thick, solid, smooth] table[x index={5}, y index={2}] {figures/data-for-paper/receive-diversity.dat};
		\addlegendentry{$\Ng=1$};
	\addplot[black, thick, dashed, smooth] table[x index={4}, y index={1}] {figures/data-for-paper/receive-diversity.dat};
	\addlegendentry{$\Ng=2$};
	\addplot[black, thick, dashdotted, smooth] table[x index={3}, y index={0}] {figures/data-for-paper/receive-diversity.dat};
		\addlegendentry{$\Ng=4$};
	\end{axis}
\end{tikzpicture}
	\caption{\changed{The rate for an arbitrary terminal when using both transmit and receive diversity. Compared to only using transmit diversity (\Figref{fig:transmit:diversity}), the extra diversity from the two antennas at the receiver brings a large gain. The relative difference between the \ostbcs shrinks, compared to the single-antenna case, because of the diminishing returns of diversity. Even with receive diversity, it is still beneficial to use transmit diversity ($\Ng=2$ or $\Ng=4$).} }\label{fig:receive:diversity}
\end{figure}

\changed{In \Figref{fig:receive:diversity}, the \CDF of the rates when using $1$, $2$ or $4$ groups and random grouping is shown. Compared to the case with a single receive antenna in \Figref{fig:transmit:diversity}, the all rates are higher with two antennas at the terminal. There is a smaller difference between the schemes since the spatial diversity, and diversity in general, suffers from diminishing returns. Still, both $\Ng=2$ and $\Ng=4$ bring gains compared to the single group case.}

\changed{In the remainder of the paper, we will continue focusing on single-antenna receivers, as we have done previously.}

\subsection{Multi-Antenna Access Points}

\changed{In order to compare the coverage performance of a cell-free massive \MIMO system to a conventional, cellular massive \MIMO system with co-located antennas, the \APs may now have $M>1$ antennas. We will assume that all antennas at a particular \AP experience the same large-scale fading (but independent small-scale fading) and that all \APs have the same number of antennas. Mathematically, this is a special case of the model in \Secref{sec:system:model} with $M\Nap$ \APs and $M$ \APs at the same location:
\[ y = \sqrt{\rho}\sum_{i=1}^{M\Nap} g_i\beta_i^{1/2}q_i + w
= \sqrt{\rho}\sum_{i=1}^{\Nap}\beta_i^{1/2} \sum_{j=1}^{M} g_{i,j}q_{i,j} + w, \]
where $g_{i,j}$ and $q_{i,j}$ are the small-scale fading and the transmitted symbol associated with antenna $j$ on \AP $i$. The grouping will be done on an antenna basis, meaning, if an \AP has several antennas, they do not necessarily belong to the same group.}

\changed{The major effects of adding more \APs in an area are twofold: the total transmit power increases and the macro diversity increases. If the output power of an \AP is the same, irrespective of the number of antennas, the total transmit power remains constant for any number of antennas at the \AP. Moreover, if all antennas at a particular \AP experience the same large-scale fading, there is no increase in macro diversity. Hence, there is no evident benefit to adding more antennas to the \APs. However, one subtle advantage to adding more antennas is due to the antenna grouping. Whenever $M\geq\Ng$, each \AP can, and should, transmit the full \ostbc, i.e., have at least one of its antennas in each group. This implies that the terminal is located equidistantly from all groups. When grouping the antennas randomly, the probability that each \AP has at least one antenna in each group increases with $M$, and with it the coverage performance. Note that the neighbor grouping ensures that each \AP transmits the full \ostbc whenever the number of antennas at each \AP exceeds the size of the \ostbc ($M\geq\Ng$), thus providing the optimal grouping. Hence, grouping the antennas in this case can increase coverage performance, albeit marginally.}

\changed{As a last example, we compare the performance of the distributed, cell-free massive \MIMO system to that of the conventional, cellular massive \MIMO system. It is difficult to get a good comparison between the two technologies, as their respective operating points and use cases may differ significantly. At any rate, to facilitate a fair comparison we let both systems have the same antenna density (same number of antennas per km$^{2}$) and the same total output power.}

\changed{To emulate a conventional massive \MIMO system, we let each \AP have $M=100$ antennas. To distinguish from the cell-free system, we further call the \AP in this case a base station. The cell-free system consists of single-antenna \APs. Since the total output power and the antenna density is constant, there will be a 100 times more \APs than base stations, but each base station will have $20$ dB more transmit power than each \AP. The cell-free system uses random grouping. The only other difference in the scenarios is that the base stations are assumed to be on a hexagonal grid.\footnote{\changed{The setting here differs from \cite{Karlsson18} in that all base stations now transmit the same system information.}} This is to emulate that base stations are deployed with more care than the \APs. As we saw from \Figref{fig:hex:vs:ppp}, all other things being equal, hexagonal deployment improves performance over \PPP deployment.}

\changed{A comparison for a system with $1000$ antennas per km$^{2}$, corresponding to a base-station distance of about $340$ meters. As expected, distributing the antennas brings a gain in the coverage performance. At the $10^{-3}$ percentile, this gain is about $6$~dB.}

\begin{figure}[ht]
\centering
\begin{tikzpicture}
	\begin{axis}[
	xlabel={rate [bpcu]},
	ylabel={\CDF},
	legend pos = {south east},
	table/col sep=comma,
	table/x index={0},
	ymode=log,
	ymax = 5e-2,
	ymin = 5e-4,
	]
	\addplot[solid] table[y index={1}, x index={3}] {figures/data-for-paper/distributed-vs-colocated.dat};
	\addlegendentry{cellular};
	\addplot[dashed] table[y index={0}, x index={2}] {figures/data-for-paper/distributed-vs-colocated.dat};
	\addlegendentry{cell-free};
	\end{axis}
\end{tikzpicture}
\caption{\changed{The \CDF of the rates for an arbitrary terminal in a cell-free and conventional massive \MIMO system with 1000 antennas per km$^{2}$. Both systems use the Alamouti code ($\Ng=2$). The base stations in the conventional case have $100$ antennas each and are placed on a hexagonal grid, while the single-antenna \APs in the cell-free case are distributed according to a \PPP. Distributing the antennas geographically over the area is beneficial leading to an increased coverage probability.}} \label{fig:distributed:vs:colocated}
\end{figure}
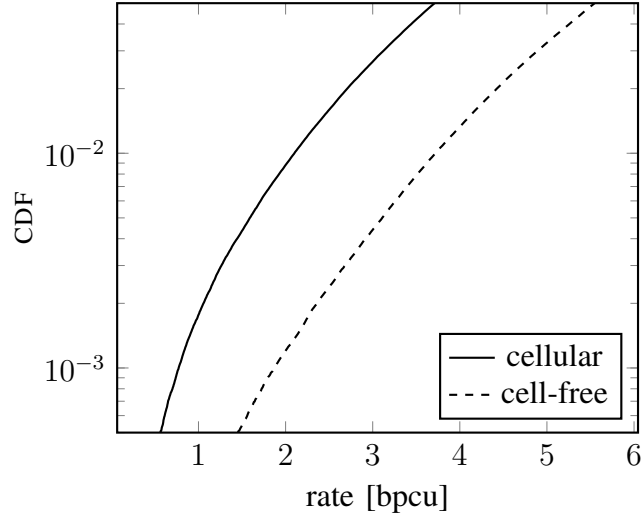

\section{Conclusions}

A cell-free massive \MIMO system will have to transmit system information to inactive terminals in order for these to join the network. This transmission needs to reach any user in the system and cannot rely on any \CSI. With downlink pilots and a space-time block code, the reliability of this transmission can be enhanced. Transmitting a space-time block code from single-antenna \APs, requires the system to divide the access points into groups. For system coverage performance, a random grouping works well, although the grouping can effect individual terminals a great deal. When multi-antenna \APs are considered, each \AP should transmit the full space-time block code if possible. Heuristically optimizing the power distribution between pilots and data can further improve the reliability.

\appendix
\subsection{Proof of \Thmref{thm:ostbc:snr}}\label{app:ostbc:snr}

In order to compute the desired \SNR in \Eqref{eq:ls:snr}, we need to compute $c_n$, $\Exp{|z_n|^{2}\given\hhat}$, and $\Exp{|\eta_n|^{2}\given\hhat}$. In some derivations, equality signs may have equation numbers above them, like $\equalref{eq:error:conditional:distribution}$, to indicate that a particular equation is useful in that step of the derivation.

First, we consider $c_n$, which is a measure of how correlated the noise $\eta_n$ is with the symbol of interest $\sn$, conditioned on $\hhat$. Note that $z_n$ is conditionally uncorrelated with the symbol $\sn$, so
\[ \Exp{\sn\conj(\eta_n + z_n)\given\hhat} = \Exp{\sn\conj\eta_n\given\hhat}. \]
The real part of $c_n$ is
\begin{equation}\label{eq:cn:real}
\begin{aligned}
\real{\Exp{\sn\conj\eta_n\given\hhat}} 
&= 
\Exp{\rsn\bar{\eta}_n\given\hhat} + \Exp{\isn\tilde{\eta}_n\given\hhat}\\
&=
-\Exp{\sqrt{\pd}\rsn\real{\hhat\herm\An\herm\Xd\be}\given\hhat} -\Exp{\sqrt{\pd}\isn\imag{\hhat\herm\Bn\herm\Xd\be}\given\hhat}\\
&\equalref{eq:error:conditional:distribution}
-\Exp{\sqrt{\pd}\rsn\real{\hhat\herm\An\herm\Xd\Ucond\hhat}\given\hhat} -\Exp{\sqrt{\pd}\isn\imag{\hhat\herm\Bn\herm\Xd\Ucond\hhat}\given\hhat}\\
&\equalref{eq:ostbcidentity:expectation:Xs}
-\dfrac{\sqrt{\pd}\Es}{2}\left(\real{\hhat\herm\An\herm\An\Ucond\hhat} +\imag{\iu\hhat\herm\Bn\herm\Bn\Ucond\hhat}\right)\\
&=
-\dfrac{\sqrt{\pd}\Es}{2}\left(\real{\hhat\herm\Ucond\hhat} +\imag{\iu\hhat\herm\Ucond\hhat}\right)\\
&=
-\dfrac{\sqrt{\pd}\Es}{2}2\real{\hhat\herm\Ucond\hhat}\\
&=
-\sqrt{\pd}\Es\hhat\herm\Ucond\hhat.
\end{aligned}
\end{equation}
In the last equality, we have used the fact that $\Ucond=\Ucond\herm$.
For the imaginary part, we calculate the cross-terms
\begin{equation}\label{eq:cn:imaginary}
\begin{aligned}
\Exp{\rsn\tilde{\eta}_1 - \isn\bar{\eta}_1\given\hhat} 
&= 
-\Exp{\rsn\imag{\sqrt{\pd}\hhat\herm\Bn\herm\Xd\htilde}
-\isn\real{\sqrt{\pd}\hhat\herm\An\herm\bX\htilde}\given\hhat}\\
&\equalref{eq:error:conditional:distribution}
-\sqrt{\pd}\Exp{\rsn\imag{\hhat\herm\Bn\herm\Xd\Ucond\hhat} 
- \isn\real{\hhat\herm\An\herm\bX\Ucond\hhat}\given\hhat}\\
&\equalref{eq:ostbcidentity:expectation:Xs}
-\dfrac{\sqrt{\pd}\Es}{2}\left(\imag{\hhat\herm\Bn\herm\An\Ucond\hhat} - \real{\iu\hhat\herm\An\herm\Bn\Ucond\hhat}\right)\\
&=
-\sqrt{\pd}\Es\imag{\hhat\herm\An\herm\Bn\Ucond\hhat}.
\end{aligned}
\end{equation}
Note that \Eqref{eq:cn:imaginary} is zero if $\An\herm\Bn$ and $\Ucond$ commute. Adding the real \Eqref{eq:cn:real} and imaginary \Eqref{eq:cn:imaginary} parts gives \Eqref{eq:cn:calculated}.

We now consider the power of the noise. The part stemming from the additive noise has power
\[ \begin{aligned}
\Exp{|z_n|^{2}\given\hhat} 
&= \Exp{|\bar{z}_n|^{2}+|\tilde{z}_n|^{2}\given\hhat}\\
&= \Exp{\Re^{2}(\hhat\herm\An\herm\bw) + \Im^{2}(\hhat\herm\Bn\herm\bw)\given\hhat}\\
&=
\dfrac{1}{2}\Re\left(\Exp{\hhat\herm\An\herm\bw\bw\trans\An\conj\hhat\conj +\hhat\herm\An\herm\bw\bw\herm\An\hhat \given\hhat}\right)\\
&+\dfrac{1}{2}\Re\left(\Exp{\hhat\herm\Bn\herm\bw\bw\trans\Bn\conj\hhat\conj +\hhat\herm\Bn\herm\bw\bw\herm\Bn\hhat \given\hhat}\right)\\
&=2\dfrac{1}{2}\hhatnorm^{2} = \hhatnorm^{2}.
\end{aligned} \]

The second noise term is more involved. We calculate the power of the real and imaginary part separately. For the real part,
\[ \begin{aligned}
&\Exp{\bar{\eta}^{2}_{n}\given \hhat} 
=
\pd\Exp{\real{\hhat\herm\An\herm\Xd\be}\real{\hhat\herm\An\herm\Xd\be}\given \hhat}\\
&=  \frac{\pd}{2}\real{\Exp{\hhat\herm\An\herm\Xd\be\be\trans\Xd\trans\An\conj\hhat\conj 
+ \hhat\herm\An\herm\Xd\be\be\herm\Xd\herm\An\hhat\given \hhat}}\\
&\equalref{eq:error:conditional:distribution}
\frac{\pd}{2}\real{\Exp{\hhat\herm\An\herm\Xd\Ucond\hhat\hhat\trans\Ucond\trans\Xd\trans\An\conj\hhat\conj\given \hhat}}\\
&+ \frac{\pd}{2}\real{\Exp{\hhat\herm\An\herm\Xd\left(\Ucond\hhat\hhat\herm\Ucond\herm + \Ccond\right)\Xd\herm\An\hhat\given \hhat}}\\
&\equalref{eq:ostbcidentity:expectation:Xs}
\frac{\pd\Es}{4}\real{\hhat\herm\An\herm
\left(\sum\limits_{k=1}^{\ns} \Ak\Ucond\hhat\hhat\trans\Ucond\trans\Ak\trans - \Bk\Ucond\hhat\hhat\trans\Ucond\trans\Bk\trans\right)\An\conj\hhat\conj}\\
&+\frac{\pd\Es}{4}\real{\hhat\herm\An\herm\left(\sum\limits_{k=1}^{\ns} \Ak\left(\Ucond\hhat\hhat\herm\Ucond\herm+\Ccond\right)\Ak\herm + \Bk\left(\Ucond\hhat\hhat\herm\Ucond\herm+\Ccond\right)\Bk\herm\right)\An\hhat}.
\end{aligned} \]
We can write this as
\begin{equation}\label{eq:noise:power:eta:real}
\Exp{\bar{\eta}^{2}_{n}\given \hhat}  
=
\dfrac{\pd\Es}{4}\left(\psi(\An,\bQ_1) + \bar{\psi}(\An,\bQ_2)\right), 
\end{equation}
where $\bQ_1$, $\bQ_2$, $\psi(\cdot,\cdot)$, and $\bar{\psi}(\cdot,\cdot)$ are defined in \Thmref{thm:ostbc:snr}.

Following the same steps, the power of the imaginary part can be written as
\begin{equation}\label{eq:noise:power:eta:imag}
\Exp{\tilde{\eta}^{2}_{n}\given \hhat}  
=
\dfrac{\pd\Es}{4}\left(\psi(\Bn,\bQ_1) - \bar{\psi}(\Bn,\bQ_2)\right). 
\end{equation}
Adding \Eqref{eq:noise:power:eta:real} and \Eqref{eq:noise:power:eta:imag} gives \Eqref{eq:noise:power:eta}.

\subsection{Proof of \Thmref{cor:snr:one:group}}\label{app:snr:one:group}
When all \APs are in the same group, they all transmit the same symbol: $s$. All variables in this proof are special cases of the variables in \Appref{app:ostbc:snr}. The variables in this section are written without a symbol index $n$, as we only consider a single symbol here ($\Ns=\Ng=1$). To be precise, $\bA=\bB=1$, $\Xd=s$, 
\[ \Ucond=\dfrac{1}{1 + \pp\tp\bar{\beta}} \] and \[ \Ccond=\dfrac{\bar{\beta}}{1 + \pp\tp\bar{\beta}}.\]

With these simplifications, we have
\[ c = -\dfrac{\sqrt{\pd}|\hat{h}|^{2}}{1 + \pp\tp\bar{\beta}}, \]
\[ \Exp{|z|^{2}\given \hat{h}} = |\hat{h}|^{2}, \]
and
\[ \Exp{|\eta|^{2}\given \hat{h}} = \pd\Es|\hat{h}|^{2}\left(\dfrac{|\hat{h}|^{2}}{(1 + \pp\tp\bar{\beta})^{2}} + \dfrac{\bar{\beta}}{1 + \pp\tp\bar{\beta}}\right). \]
Thus, the \SNR is given by
\[ \snr^{\LS} \triangleq \dfrac{\pd\Es|\hat{h}|^{2}}{1 + \dfrac{\pd\Es\bar{\beta}}{\pp\tp\bar{\beta} + 1}}\left(1 - \dfrac{1}{1+\pp\tp\bar{\beta}}\right)^{2} 
=
\dfrac{\pd\Es|\hat{h}|^{2}}{1 + \dfrac{\pd\Es\bar{\beta}}{\pp\tp\bar{\beta} + 1}}\left(\dfrac{\pp\tp\bar{\beta}}{1+\pp\tp\bar{\beta}}\right)^{2}  \]
and since 
\[ |\hat{h}|^{2}\sim \dfrac{\pp\tp\bar{\beta} + 1}{2\pp\tp}\chi^{2}_2 \sim \dfrac{\pp\tp\bar{\beta} + 1}{2\pp\tp}\ExpDist{1/2}\sim\ExpDist{\dfrac{\pp\tp}{\pp\tp\bar{\beta}+1}},\]
we have the desired result.
\bibliographystyle{IEEEtran} 
\bibliography{IEEEabrv,cf-system-info} 

\begin{thebibliography}{10}
\providecommand{\url}[1]{#1}
\csname url@samestyle\endcsname
\providecommand{\newblock}{\relax}
\providecommand{\bibinfo}[2]{#2}
\providecommand{\BIBentrySTDinterwordspacing}{\spaceskip=0pt\relax}
\providecommand{\BIBentryALTinterwordstretchfactor}{4}
\providecommand{\BIBentryALTinterwordspacing}{\spaceskip=\fontdimen2\font plus
\BIBentryALTinterwordstretchfactor\fontdimen3\font minus
  \fontdimen4\font\relax}
\providecommand{\BIBforeignlanguage}[2]{{%
\expandafter\ifx\csname l@#1\endcsname\relax
\typeout{** WARNING: IEEEtran.bst: No hyphenation pattern has been}%
\typeout{** loaded for the language `#1'. Using the pattern for}%
\typeout{** the default language instead.}%
\else
\language=\csname l@#1\endcsname
\fi
#2}}
\providecommand{\BIBdecl}{\relax}
\BIBdecl

\bibitem{Nayebi15}
E.~Nayebi, A.~Ashikhmin, T.~L. Marzetta, and H.~Yang, ``Cell-free massive
  {{MIMO}} systems,'' in \emph{2015 49th {{Asilomar Conference}} on
  {{Signals}}, {{Systems}} and {{Computers}}}, Nov. 2015, pp. 695--699.

\bibitem{Ngo17a}
H.~Q. Ngo, A.~Ashikhmin, H.~Yang, E.~G. Larsson, and T.~L. Marzetta,
  ``Cell-free massive {{MIMO}} versus small cells,'' \emph{IEEE Transactions on
  Wireless Communications}, vol.~16, no.~3, pp. 1834--1850, Mar. 2017.

\bibitem{Nayebi17}
E.~Nayebi, A.~Ashikhmin, T.~L. Marzetta, H.~Yang, and B.~D. Rao, ``Precoding
  and {{Power Optimization}} in {{Cell}}-{{Free Massive MIMO Systems}},''
  \emph{IEEE Transactions on Wireless Communications}, vol.~16, no.~7, pp.
  4445--4459, Jul. 2017.

\bibitem{Nguyen17}
L.~D. Nguyen, T.~Q. Duong, H.~Q. Ngo, and K.~Tourki, ``Energy {{Efficiency}} in
  {{Cell}}-{{Free Massive MIMO}} with {{Zero}}-{{Forcing Precoding Design}},''
  \emph{IEEE Communications Letters}, vol.~21, no.~8, pp. 1871--1874, Aug.
  2017.

\bibitem{Ngo18}
H.~Q. Ngo, L.~N. Tran, T.~Q. Duong, M.~Matthaiou, and E.~G. Larsson, ``On the
  {{Total Energy Efficiency}} of {{Cell}}-{{Free Massive MIMO}},'' \emph{IEEE
  Transactions on Green Communications and Networking}, vol.~2, no.~1, pp.
  25--39, Mar. 2018.

\bibitem{Interdonato18}
G.~Interdonato, E.~Bj\"ornson, H.~Q. Ngo, P.~Frenger, and E.~G. Larsson,
  ``Ubiquitous {{Cell}}-{{Free Massive MIMO Communications}},''
  \emph{arXiv:1804.03421 [cs, math]}, Apr. 2018.

\bibitem{Bashar18}
M.~Bashar, K.~Cumanan, A.~G. Burr, H.~Q. Ngo, and H.~V. Poor, ``Mixed
  {{Quality}} of {{Service}} in {{Cell}}-{{Free Massive MIMO}},'' \emph{IEEE
  Communications Letters}, pp. 1--1, 2018.

\bibitem{Marzetta16}
T.~L. Marzetta, E.~G. Larsson, H.~Yang, and H.~Q. Ngo, \emph{Fundamentals of
  Massive {{MIMO}}}.\hskip 1em plus 0.5em minus 0.4em\relax Cambridge:
  {Cambridge University Press}, 2016.

\bibitem{Bjornson17}
E.~Bj\"ornson, J.~Hoydis, and L.~Sanguinetti, ``Massive {{MIMO}} networks:
  Spectral, energy, and hardware efficiency,'' \emph{Foundations and
  Trends\textregistered{} in Signal Processing}, vol.~11, no. 3-4, pp.
  154--655, 2017.

\bibitem{Irmer11}
R.~Irmer, H.~Droste, P.~Marsch, M.~Grieger, G.~Fettweis, S.~Brueck, H.~P.
  Mayer, L.~Thiele, and V.~Jungnickel, ``Coordinated multipoint: {{Concepts}},
  performance, and field trial results,'' \emph{IEEE Communications Magazine},
  vol.~49, no.~2, pp. 102--111, Feb. 2011.

\bibitem{Shamai01}
S.~Shamai and B.~M. Zaidel, ``Enhancing the cellular downlink capacity via
  co-processing at the transmitting end,'' in \emph{{{IEEE VTS}} 53rd
  {{Vehicular Technology Conference}}}, Rhodes, 2001, pp. 1745--1749.

\bibitem{Gesbert10}
D.~Gesbert, S.~Hanly, H.~Huang, S.~S. Shitz, O.~Simeone, and W.~Yu,
  ``Multi-cell {{MIMO}} cooperative networks: {{A}} new look at
  {{Interference}},'' \emph{IEEE Journal on Selected Areas in Communications},
  vol.~28, no.~9, pp. 1380--1408, Dec. 2010.

\bibitem{Zhou03}
S.~Zhou, M.~Zhao, X.~Xu, J.~Wang, and Y.~Yao, ``Distributed wireless
  communication system: A new architecture for future public wireless access,''
  \emph{IEEE Communications Magazine}, vol.~41, no.~3, pp. 108--113, Mar. 2003.

\bibitem{Bjornson10}
E.~Bj\"ornson, R.~Zakhour, D.~Gesbert, and B.~Ottersten, ``Cooperative
  multicell precoding: Rate region characterization and distributed strategies
  with instantaneous and statistical {{CSI}},'' \emph{IEEE Transactions on
  Signal Processing}, vol.~58, no.~8, pp. 4298--4310, Aug. 2010.

\bibitem{Buzzi17}
S.~Buzzi and C.~D'Andrea, ``Cell-free massive {{MIMO}}: User-centric
  approach,'' \emph{IEEE Wireless Communications Letters}, vol.~6, no.~6, pp.
  706--709, Dec. 2017.

\bibitem{Lozano13}
A.~Lozano, R.~W. Heath, and J.~G. Andrews, ``Fundamental limits of
  cooperation,'' \emph{IEEE Transactions on Information Theory}, vol.~59,
  no.~9, pp. 5213--5226, Sep. 2013.

\bibitem{Chen18}
Z.~Chen and E.~Bj\"ornson, ``Channel hardening and favorable propagation in
  cell-free massive {{MIMO}} with stochastic geometry,'' \emph{IEEE
  Transactions on Communications}, pp. 1--1, 2018.

\bibitem{Dahlman11}
E.~Dahlman, S.~Parkvall, and J.~Sk\"old, \emph{{{4G}}:
  {{LTE}}/{{LTE}}-{{Advanced}} for Mobile Broadband}, 1st~ed.\hskip 1em plus
  0.5em minus 0.4em\relax {Academic Press}, 2011.

\bibitem{Karlsson18}
M.~Karlsson, E.~Bj\"ornson, and E.~G. Larsson, ``Performance of in-band
  transmission of system information in massive {{MIMO}} systems,'' \emph{IEEE
  Transactions on Wireless Communications}, vol.~17, no.~3, pp. 1700--1712,
  Mar. 2018.

\bibitem{Meng16}
X.~Meng, X.~Gao, and X.~G. Xia, ``Omnidirectional precoding based transmission
  in massive {{MIMO}} systems,'' \emph{IEEE Transactions on Communications},
  vol.~64, no.~1, pp. 174--186, Jan. 2016.

\bibitem{Andrews11}
J.~G. Andrews, F.~Baccelli, and R.~K. Ganti, ``A tractable approach to coverage
  and rate in cellular networks,'' \emph{IEEE Transactions on Communications},
  vol.~59, no.~11, pp. 3122--3134, Nov. 2011.

\bibitem{Lu15}
W.~Lu and M.~Di~Renzo, ``Stochastic geometry modeling of cellular networks:
  {{Analysis}}, simulation and experimental validation,'' in \emph{Proceedings
  of the 18th {{ACM International Conference}} on {{Modeling}}, {{Analysis}}
  and {{Simulation}} of {{Wireless}} and {{Mobile Systems}}}.\hskip 1em plus
  0.5em minus 0.4em\relax New York, NY, USA: {ACM}, 2015, pp. 179--188.

\bibitem{ElSawy17}
H.~ElSawy, A.~{Sultan-Salem}, M.~S. Alouini, and M.~Z. Win, ``Modeling and
  analysis of cellular networks using stochastic geometry: {{A}} tutorial,''
  \emph{IEEE Communications Surveys Tutorials}, vol.~19, no.~1, pp. 167--203,
  2017.

\bibitem{Weber12}
S.~Weber and J.~G. Andrews, ``\BIBforeignlanguage{English}{Transmission
  capacity of wireless networks},''
  \emph{\BIBforeignlanguage{English}{Foundations and Trends\textregistered{} in
  Networking}}, vol.~5, no. 2\textendash{}3, pp. 109--281, 2012.

\bibitem{Larsson03}
E.~G. Larsson and P.~Stoica, \emph{Space-Time Block Coding for Wireless
  Communications}.\hskip 1em plus 0.5em minus 0.4em\relax Cambridge: {Cambridge
  University Press}, 2003.

\bibitem{Medard00}
M.~Medard, ``The effect upon channel capacity in wireless communications of
  perfect and imperfect knowledge of the channel,'' \emph{IEEE Transactions on
  Information Theory}, vol.~46, no.~3, pp. 933--946, May 2000.

\bibitem{Kay93}
S.~M. Kay, \emph{\BIBforeignlanguage{English}{Fundamentals of Statistical
  Signal Processing, Volume {{I}}: Estimation Theory}}, 1st~ed.\hskip 1em plus
  0.5em minus 0.4em\relax Englewood Cliffs, N.J: {Prentice Hall}, Apr. 1993.

\bibitem{Bjornson09}
E.~Bj\"ornson, D.~Hammarwall, and B.~Ottersten, ``Exploiting quantized channel
  norm feedback through conditional statistics in arbitrarily correlated
  {{MIMO}} systems,'' \emph{IEEE Transactions on Signal Processing}, vol.~57,
  no.~10, pp. 4027--4041, Oct. 2009.

\bibitem{Rappaport01}
T.~Rappaport, \emph{Wireless Communications: {{Principles}} and Practice},
  2nd~ed.\hskip 1em plus 0.5em minus 0.4em\relax Upper Saddle River, NJ, USA:
  {Prentice Hall PTR}, 2001.

\bibitem{Gudmundson91}
M.~Gudmundson, ``Correlation model for shadow fading in mobile radio systems,''
  \emph{Electronics Letters}, vol.~27, no.~23, pp. 2145--2146, Nov. 1991.

\bibitem{Cheng17}
H.~V. Cheng, E.~Bj\"ornson, and E.~Larsson, ``Optimal pilot and payload power
  control in single-cell massive {{MIMO}} systems,'' \emph{IEEE Transactions on
  Signal Processing}, vol.~65, no.~9, pp. 2363--2378, May 2017.

\bibitem{Alamouti98}
S.~M. Alamouti, ``A simple transmit diversity technique for wireless
  communications,'' \emph{IEEE Journal on Selected Areas in Communications},
  vol.~16, no.~8, pp. 1451--1458, Oct. 1998.

\end{thebibliography}
\end{document}